\documentclass[aps,prl,reprint,twocolumn,numerical,amsmath,amssymb,a4paper,graphicx,superscriptaddress,longbibliography,noshowpacs]{revtex4-1}

\usepackage[utf8]{inputenc}
\usepackage[english]{babel}
\usepackage[
	pdffitwindow=true,
	colorlinks=true,
	frenchlinks=false,
        linkcolor=violet,
	anchorcolor=violet,
        citecolor=violet,
        filecolor=violet,
        urlcolor=violet,
        bookmarks=true,
        bookmarksopen=true,
	bookmarksnumbered=true, 
        bookmarksopenlevel=1,
        plainpages=false,
	pdfpagelayout=TwoPageLeft,
        pdfpagelabels=true,
	breaklinks
]{hyperref}
\usepackage{graphicx}
\usepackage{siunitx}
\usepackage{xcolor}
\usepackage{ulem}
\DeclareUnicodeCharacter{2212}{-}

\newcommand{\gammas}{\gamma_s}
\newcommand{\kappac}{\kappa_c}
\newcommand{\kappaext}{\kappa_\mathrm{ext}}
\newcommand{\geff}{g_\mathrm{eff}}
\newcommand{\omegac}{\omega_c}

\begin{document}

\title{Echo trains in pulsed electron spin resonance of a strongly coupled spin ensemble}

\author{Stefan Weichselbaumer}
\affiliation{Walther-Mei{\ss}ner-Institut, Bayerische Akademie der Wissenschaften, 85748 Garching, Germany }
\affiliation{Physik-Department, Technische Universit\"{a}t M\"{u}nchen, 85748 Garching, Germany}

\author{Matthias Zens}
\affiliation{Institute for Theoretical Physics, TU Wien, Wiedner Hauptstraße 8-10/136, 1040 Vienna, Austria}
\affiliation{ITAMP, Harvard-Smithsonian Center for Astrophysics, Cambridge, Massachusetts 02138, USA }

\author{Christoph W.\ Zollitsch}
\altaffiliation[Present address: ]{London Centre for Nanotechnology, University College London, London WC1H 0AH, United Kingdom}
\affiliation{Walther-Mei{\ss}ner-Institut, Bayerische Akademie der Wissenschaften, 85748 Garching, Germany }
\affiliation{Physik-Department, Technische Universit\"{a}t M\"{u}nchen, 85748 Garching, Germany}

\author{Martin S.\ Brandt}
\affiliation{Walter Schottky Institut, Technische Universit\"{a}t M\"{u}nchen, 85748 Garching, Germany}
\affiliation{Physik-Department, Technische Universit\"{a}t M\"{u}nchen, 85748 Garching, Germany}
\affiliation{Munich Center for Quantum Science and Technology (MCQST), Schellingstra{\ss}e 4, 80799 M\"{u}nchen, Germany}

\author{Stefan Rotter}
\affiliation{Institute for Theoretical Physics, TU Wien, Wiedner Hauptstraße 8-10/136, 1040 Vienna, Austria}

\author{Rudolf Gross}
\affiliation{Walther-Mei{\ss}ner-Institut, Bayerische Akademie der Wissenschaften, 85748 Garching, Germany }
\affiliation{Physik-Department, Technische Universit\"{a}t M\"{u}nchen, 85748 Garching, Germany}
\affiliation{Munich Center for Quantum Science and Technology (MCQST), Schellingstra{\ss}e 4, 80799 M\"{u}nchen, Germany}

\author{Hans Huebl}
\email[]{hans.huebl@wmi.badw.de}
\affiliation{Walther-Mei{\ss}ner-Institut, Bayerische Akademie der Wissenschaften, 85748 Garching, Germany }
\affiliation{Physik-Department, Technische Universit\"{a}t M\"{u}nchen, 85748 Garching, Germany}
\affiliation{Munich Center for Quantum Science and Technology (MCQST), Schellingstra{\ss}e 4, 80799 M\"{u}nchen, Germany}

\date{\today}

\begin{abstract}
    We report on a novel dynamical phenomenon in electron spin resonance experiments of
    phosphorus donors.  When strongly coupling the paramagnetic ensemble to a superconducting lumped element resonator, the coherent exchange between these two subsystems leads to a train of periodic, self-stimulated echos after a conventional Hahn echo pulse sequence. The presence of these multi-echo signatures is explained using a simple model based on spins rotating on the Bloch sphere, backed up by numerical calculations using the inhomogeneous Tavis-Cummings Hamiltonian.    
\end{abstract}


\maketitle
\begin{figure}
    \centering
    \includegraphics[width=3.3in]{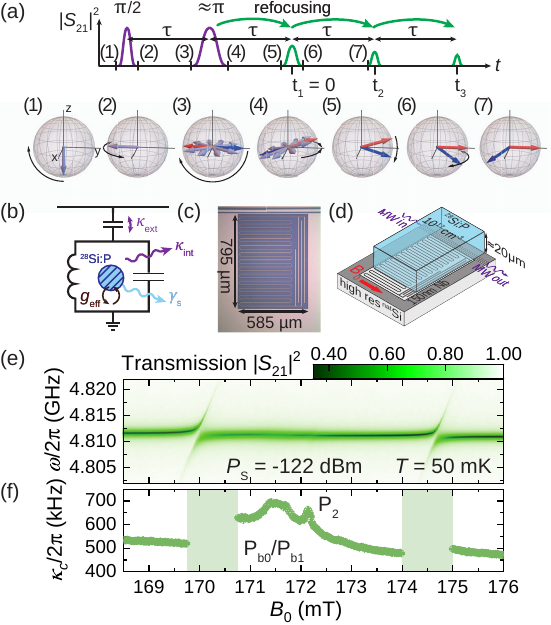}%
    \caption{\label{Fig1}(a)~Schematic of the Hahn echo sequence and the associated
        states in the Bloch sphere (exterior black arrows indicate the ensuing spin dynamics). A $\pi/2$-pulse is applied between (1) and (2) and an imperfect $\pi$-pulse between (3) and (4), leading to the first (conventional) Hahn echo between (5) and (6). For the subsequent pulse train observed in the strong coupling limit the spin packets indicated by blue and red arrows are crucial, which lie in opposite $S_y$-directions when the first $\pi$ pulse arrives at (3) (see text). In panels (5)-(7) only these two spin packets are shown for better visibility.
    (b)~Circuit diagram and 
    (c)~microscope image of the lumped element resonator (LER). 
    (d)~Schematic of the \SI{20}{\micro\meter} thin $^{28}$Si:P sample mounted
    on top of the LER  also indicating the in-plane magnetic field direction $B_0$. 
    (e)~Normalized transmission $\left|S_{21}\right|^2$ as a function of frequency and
    magnetic field. 
    Two avoided crossings are visible, indicating strong
    coupling between the hyperfine-split transitions of the phosphorus donors and the
    resonator. (f)~Extracted linewidth $\kappa/2\pi$ (HWHM) as a function of
    magnetic field. Two additional spectroscopic features are observed, which are
    attributed to dangling bond defects P\textsubscript{b0}/P\textsubscript{b1}  and P$_2$ dimers, respectively (see text).}%
\end{figure}

Pulsed electron spin resonance (ESR) is an essential spectroscopy technique used
in many fields of science, e.g., for the study of the structure and dynamics of
molecular systems\,\cite{Prisner2001,Eaton2015}, for material
science\,\cite{Baranov2017} as well as for quantum sensing and information
applications\,\cite{Schirhagl2014,Devoret2013,Zwanenburg2013}. To implement this
technique, a vast repertoire of sophisticated pulse sequences
exists\,\cite{Schweiger2001}, each of them optimized to investigate particular
spin properties. Nevertheless, the majority of the sequences is based on a Hahn
echo\,\cite{Hahn1950} as schematically shown in Fig.\,\ref{Fig1}\,(a).

A newly emerging area for ESR experiments is the processing of quantum information.
Using superconducting microwave resonators, the so-called ``strong coupling
regime'' has recently been demonstrated\,\cite{Kubo2010, Schuster2010, Amsuss2011, Probst2013, Putz2014, Zollitsch2015}. 
Here, the coherent exchange of information between the microwave resonator and the
spin ensemble exceeds the individual decay rates of the two subsystems, which is
a requirement for applications involving the storage and conversion of quantum information\,\cite{Morton2008,Kubo2010,Bushev2011,Grezes2016}.
Apart from its importance for quantum technology, a strong coupling rate also enhances the sensitivity in ESR
applications\,\cite{Bienfait2016,Eichler2017} going beyond classical ESR models\,\cite{Schweiger2001,Levitt2008}. First seminal experiments in the presence of strong spin-photon coupling  revealed a plethora of new physical effects\,\cite{Putz2014,Rose2017,Putz2017a,Angerer2017}. 
A fascinating question that remains unresolved, however, is what happens when the Hahn echo is transferred to the context of a strongly coupled spin ensemble.

To explore this question experimentally, we work with  a superconducting microwave resonator strongly coupled to a paramagnetic spin ensemble.
Specifically, we compare pulsed ESR measurements of a strongly coupled spin
ensemble based on isolated phosphorus donors in a $^{28}$Si host matrix with a
weakly coupled ensemble of P$_2$ dimers also present in the sample. In the weak coupling case, as in a conventional ESR experiment, we observe a single Hahn echo in terms of a photon pulse that is emitted into the resonator at $2\tau$ when the spins refocus, where $\tau$ is the inter-pulse delay. In stark contrast, when applying the same Hahn echo sequence in the strong coupling regime, we observe a periodic sequence of spin echo signatures spaced by $\tau$. Although this phenomenon has been reported for up to two echos earlier\,\cite{Gordon1958}, it was not set in context with the strong coupling regime and a thorough understanding of the underlying mechanism is missing. Here we show that the formation of self-stimulated echos is a robust phenomenon and can be well understood based on the inhomogeneous Tavis-Cummings model. 

Our experimental scheme is shown in Fig.~\ref{Fig1}\,(b)--(d) and consists of
a planar su\-per\-con\-ducting lumped element resonator (LER), which is patterned into a \SI{150}{nm} thin Nb film
on an intrinsic $^\mathrm{nat}$Si substrate \cite{SupplementaryInformation}. The LER is located next to a microwave feedline allowing us to probe the complex microwave transmission of the device. A \SI{20}{\micro\meter} thin slab of $[100]$ oriented $^{28}$Si:P is mounted onto the LER (see Fig.~\ref{Fig1}\,(d)) and investigated at a temperature of $T = \SI{50}{mK}$. A static magnetic field $B_0$ is applied parallel to the Nb film to avoid degradation of its superconducting
properties.  We perform continuous-wave (cw) ESR by measuring the microwave transmission $\left|S_{21}\right|^2$ of the chip using a vector network analyzer. For pulsed ESR experiments, we digitize the echo signal using a heterodyne down-conversion scheme\,\cite{SupplementaryInformation}.

\textit{Continuous-wave ESR spectroscopy. }We first perform
cw ESR spectroscopy to pre-characterize the sample. Figure~\ref{Fig1}\,(e)
shows the normalized microwave transmission $\left|S_{21}\right|^2$ for an incident power on the sample of $P_\mathrm{S} = \SI{-122}{dBm}$.
At $B_0 = \SI{168.5}{mT}$, we observe a bare resonator frequency of
$\omegac/2\pi = \SI{4.8116}{GHz}$. Using a robust circle-fitting algorithm\,\cite{Probst2015}, we determine a
half-width-at-half-maximum (HWHM) line width of $\kappac/2\pi = \SI{534.85}{kHz}$, corresponding to a total quality factor of $Q = \omegac/2\kappac = 4498$. The coupling rate of the resonator to the feedline is $\kappa_\mathrm{ext}/2\pi = \SI{304.15}{kHz}$. Similarly, we extract the spin relaxation rate using a Lorentzian fit along the field axis far detuned from the resonator and find $\gammas/2\pi = \SI{279.03}{kHz}$. We observe two distinct avoided crossings at
$B_0 = \SI{170.1}{\milli\tesla}$ and $B_0 = \SI{174.3}{\milli\tesla}$, which are associated with  the two hyperfine-split lines of phosphorus donors in silicon. 
The presence of the avoided crossings suggests that the spin ensembles of these
isolated phosphorus donors couple strongly to the LER. 
We determine the corresponding coupling rate $\geff/2\pi =\SI{1.54}{MHz}$ from the vaccum Rabi splitting at $B_0 = 170.19\,\mathrm{mT}$,  corresponding to a cooperativity $C =\geff^2/(\kappac\gammas) = 12.2$.  Note that the single spin-resonator coupling rate is not spatially uniform\,\cite{Weichselbaumer2019, SupplementaryInformation}.

We obtain information about further spin species present in the sample by analyzing the resonator linewidth $\kappac$ as a function of the magnetic field outside the avoided crossings from the data in Fig.\,\ref{Fig1}\,(e). We find in  Fig.\,\ref{Fig1}\,(f) 
a broad structure at $B_0 = \SI{171.5}{\milli\tesla}$, which is assigned to  dangling bond defects at the (100)Si/SiO$_2$ interface, also known as P$_\mathrm{b0}$/P$_\mathrm{b1}$ defects\,\cite{Poindexter1981,Stesmans1998}, and a sharp signature at $B_0 = \SI{172.2}{\milli\tesla}$ corresponding to statistically formed exchange-coupled donor pairs, called P$_2$ dimers, with a concentration $\left[\mathrm{P}_2\right]\ll\left[\mathrm{P}\right]$\,\cite{Feher1955,Jerome1964,Morigaki1972,Shankar2015}.
The analysis of this  P$_2$ dimer peak \cite{Herskind2009} yields a spin relaxation rate $\gamma_\mathrm{s,P_2}/2\pi=\SI{1.74}{MHz}$ and an effective coupling rate $g_\mathrm{eff,P_2}/2\pi = \SI{0.35}{MHz}$. This sets the $\mathrm{P}_2$ dimers in the weak coupling regime with $C = 0.13$, as  expected from the $\sqrt{N}$ scaling of $\geff$  \cite{Huebl2013,Zollitsch2015}. Hence, we can use these two spin ensembles to directly compare the dynamics in the  weak and strong coupling regime under the same experimental conditions.

\begin{figure}
    \centering
    \includegraphics[width=3.3in]{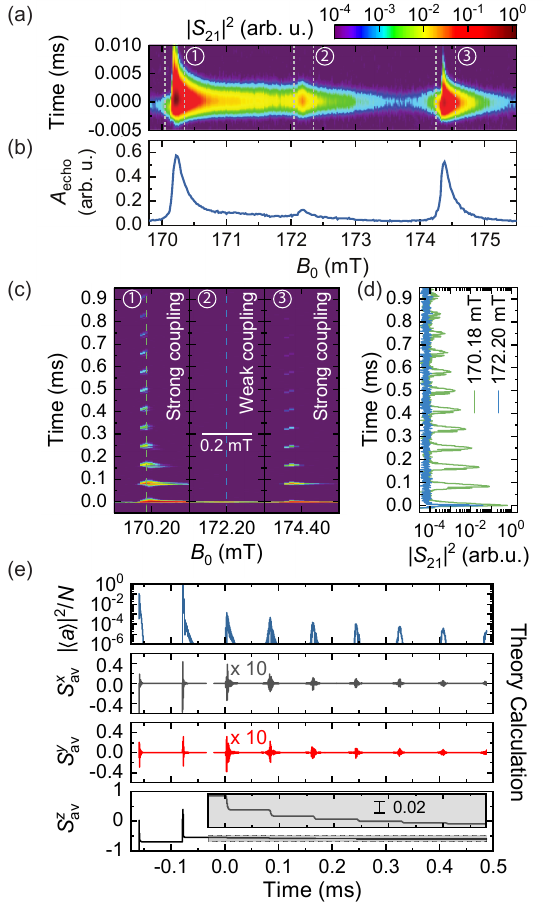}%
    \caption{\label{Fig2}(a)  Echo signal as a function of acquisition time and magnetic field of the first echo. Dashed lines indicate the sub-panels in (c).
        (b) Integrated echo area $A_\mathrm{echo}=\int_\mathrm{echo}|S_{21}|\mathrm{d}t$ as a function of magnetic field showing the  two hyperfine transitions as well as the P$_2$ dimer peak. 
        (c) Microwave signal intensity $|S_{21}|^2$ displaying several echo signals after the
        conventional first echo for the strongly coupled hyperfine lines ($\raisebox{.5pt}{\textcircled{\raisebox{-.9pt} {1}}}$ and $\raisebox{.5pt}{\textcircled{\raisebox{-.9pt} {3}}}$), while only one echo is visible for weakly coupled P$_2$
        dimers ($\raisebox{.5pt}{\textcircled{\raisebox{-.9pt} {2}}}$).
        (d) Microwave signal intensity $|S_{21}|^2$ for fixed magnetic field [cf. dashed lines in
        (c)] for the hyperfine transition (green) and P$_2$ line (blue). 
        (e) Temporal evolution of the average resonator photon number $|\langle a\rangle|^2/N$ (upper panel), the average spin expectation values $S_\text{av}^{x,y}=\sum_j\langle\sigma_j^{x,y}\rangle/N$ (middle panels), and $S_\text{av}^{z}=\sum_j\langle\sigma_j^{z}\rangle/N$ (lower panel) calculated from the semiclassical Maxwell-Bloch equations. The inset shows a zoom of the gray shaded area.}%
\end{figure}

\textit{Pulsed ESR spectroscopy. }In a next step we now apply a Hahn-type echo sequence based on two Gaussian-shaped pulses with a width of \SI{1}{\micro\second} and \SI{2}{\micro\second}, and a pulse spacing of $\tau = \SI{80}{\mu s}$.  We use a  fixed frequency $\omega_\text{p}/2\pi = \SI{4.8116}{GHz}$, even though $\omegac$ slightly shifts with $B_0$ [see Fig.~\ref{Fig1}\,(e)].
Figure~\ref{Fig2}\,(a) shows the Hahn echo-detected field sweep of the first echo in the time domain, where we have set the origin of the time axis to the
maximum of this first echo.
Note that all data shown here are single-shot measurements  and no
signal averaging is performed. The time interval between measurements at subsequent field points is \SI{300}{s}, chosen to be long compared to the spin relaxation time $T_1 = \SI{32.4\pm0.8}{s}$\,\cite{SupplementaryInformation}. From an analysis of the collective coupling rate we estimate the absolute number of spins addressed in the spin echo to be $\approx\num{1.06e10}$\,\cite{SupplementaryInformation}. For the spin-sensitivity, we obtain $\approx\SI{1.15e5}{spins/\sqrt{Hz}}$ assuming a  repetition time of $5\,T_1$ and a signal-to-noise ratio of one \cite{Eichler2017,Bienfait2016}.

Figure~\ref{Fig2}\,(b) displays the echo area $A_\mathrm{echo}=\int_\mathrm{echo}|S_{21}| \mathrm{d}t$ using the data from panel~(a)
showing three peaks corresponding to the 
hyperfine as well as the P$_2$ dimer transition. Evidently, the line widths of the hyperfine transitions are much wider than expected from  $\gammas$. Due to the presence of strong coupling, the spin system hybridizes such that the line width should also reflect  $\geff$. Additionally, both peaks are asymmetric with a tail towards larger magnetic fields, which we attribute to the excitation with a fixed $\omega_\mathrm{p}$  without compensating for the dispersion of the avoided crossing. Moreover, the absence of a clear echo corresponding to the
P$_\mathrm{b0}$/P$_\mathrm{b1}$ defects is related to their small   $T_2=\SI{22}{\micro\second}$\,\cite{SupplementaryInformation}.

The conclusion we draw from this analysis is that one observes the first \textit{conventional} Hahn echo (at $t = 0$) for both, the weakly and the strongly coupled spin ensembles. The fundamental difference between weak and strong coupling  manifests itself only on longer time scales as shown in Fig.~\ref{Fig2}\,(c). Particular attention deserve subpanels $\raisebox{.5pt}{\textcircled{\raisebox{-.9pt} {1}}}$ and $\raisebox{.5pt}{\textcircled{\raisebox{-.9pt} {3}}}$, corresponding to the strong coupling case. Here, the first Hahn echo is followed by a periodic sequence of echo signatures, which are timed with a delay equal to the pulse delay $\tau$.
In contrast, only the first conventional Hahn echo is present for the weakly coupled P$_2$ dimers shown in subpanel $\raisebox{.5pt}{\textcircled{\raisebox{-.9pt} {2}}}$\,\footnote{Note that we use  for the P$_2$ dimer measurements the same measurement protocol as for  hyperfine transitions. Within the noise budget, a second echo should be detectable, if its echo amplitudes scale in the same manner as for the strong coupling case. However, we do not observe such a subsequent echo.}.

This marked difference is even more apparent in Fig.~\ref{Fig2}\,(d), where we show time traces recorded at the fixed magnetic fields of $B_0 =
172.20\,\mathrm{mT}$ and $170.18\,\mathrm{mT}$ [dashed lines in Fig.~\ref{Fig2}\,(c)] corresponding to the weak and strong coupling regime. While only the first conventional Hahn echo appears for the P$_2$ dimers ($T_{2,P_2} = \SI{4.67\pm 0.13}{ms}$\,\cite{SupplementaryInformation}), we observe 12 echos separated
by $\tau$ for the strongly coupled hyperfine transitions ($T_\mathrm{2,P} =
\SI{2.37\pm0.08}{ms}$\,\cite{SupplementaryInformation}). 
The echo signatures in the echo train exhibit an underlying substructure going beyond the scope of this manuscript.
Although several mechanisms of multiple echo generation are known in the literature, the absence of multiple echos for the P$_2$ dimers exclude these for a possible explanation (see also Ref.\,\cite{SupplementaryInformation} for more details). This suggests that the detection of the multiple echos is, indeed, related to the strong coupling regime. 

The relevant mechanism leading to this unique dynamical evolution can be best understood when revisiting the conventional Hahn echo sequence shown in Fig.\,\ref{Fig1}\,(a). For simplicity, we assume here that all spins end up in the $xy$-plane after the first $\pi/2$-pulse (see panels 1-3), although the spatial variation of the excitation field $B_1$ and the frequency distribution of the spin ensemble inevitably lead to rotation errors. Realistically, the net dipole moment generated in the $xy$-plane during this first pulse leads to a strong collective coupling with the resonator and, hence, rapid deexcitation of the spin system. However, dephasing quickly reduces this dipole moment and thereby effectively suppresses this spin decay channel. After an evolution time $\tau$, the second pulse is injected to start the refocusing process. A perfect $\pi$ pulse would lead to a refocusing of all spins after another time span $\tau$, creating the first (conventional) Hahn echo without any subsequent echos. With the rotation angle realistically deviating from $\pi$, however, the refocusing is imperfect and the spins end up at different latitudes on the Bloch sphere, depending on their detuning $\delta\omega$ from the average Larmor frequency (see panels 3-5). This mechanism can also be understood as a frequency encoding of spin packets depending on their orientation on the Bloch sphere at the arrival time of this imperfect $\pi$-pulse. Specifically, we identify spin packets that point in opposite directions on the $S_y$-axis when the imperfect $\pi$-pulse arrives using red and blue arrows in the panels in Fig.\,\ref{Fig1}\,(a). These will be particularly relevant for the subsequent pulse train. Their frequency detunings are determined by those multiples of $\pi$-rotations that the spins already undertook at the arrival of the refocusing pulse: $\delta\omega=2n\pi/\tau$ (red spins) and $\delta\omega= (2n+1)\pi/\tau$ (blue spins) with $n\in \mathbb{Z}$. In this way, spins with significantly different individual detuning values $\delta\omega$ are now encoded in the same packet. At a time $\tau$ after the imperfect $\pi$-pulse, when spins (partially) refocus, they emit the first (conventional) Hahn echo through the coupling to the resonator. Notably, the net dipole moment in the $xy$-plane created in this refocusing process, together with the strong coupling to the microwave field also leads to a significant spin decay. Importantly, this decay is realized on the Bloch sphere as a spin rotation during this first Hahn echo that affects the projection of the dipole moment on the $xy$-plane differently for the blue and red spin packets (see panels 5 \& 6). This rotation becomes significant at a time $\tau$ after the first Hahn echo, where these spin packets again point in opposite $S_y$-directions (red and blue arrows in panel 7): with the $xy$ projection of these two spin vectors now having different lengths, they give rise to another net dipole moment that produces the (unconventional) second Hahn echo. Here, the process starts all over again, producing the third echo etc.

Note that without the spin rotation during the first Hahn echo, the red and blue spins would maintain the same $xy$ projection, such that no net dipole is created and therefore also no subsequent echos. In this way, one not only understands how the generation of one echo gives rise to the next one, but also why the strong coupling regime is essential: for weak coupling also the spin rotation by deexcitation through the resonator is weak, such that all unconventional echos are negligibly small. Moreover, also imperfect rotation angles are essential (as induced, e.g., by the inhomogeneities in the system), as no frequency encoding of spin packets would occur otherwise (for more information see  \cite{SupplementaryInformation}). 

\textit{Theoretical description. }
To underpin this heuristic explanation, we set up a theoretical model based on the inhomogeneous Tavis-Cummings Hamiltonian,
\begin{eqnarray}
\mathcal{H}=&&\hbar\Delta_c\,a^\dag a +\frac{\hbar}{2}\sum_{j=1}^N\Delta_j \sigma_j^z
+\sum_{j=1}^N\hbar[g_j\sigma_j^-a^\dag+g_j^*\sigma_j^+a]\nonumber\\
&&+i\hbar[\eta(t) a^\dag-\eta^*(t) a],
\label{eq:Hamiltonian}
\end{eqnarray}
 where $\Delta_c\equiv \omega_c-\omega_p$ and $\Delta_j\equiv\omega_j-\omega_p$ are the detunings of the resonator frequency $\omega_c$ and of the individual spin frequencies $\omega_j$ from the carrier frequency $\omega_p$ of the incoming microwave pulse with amplitude $\eta(t)$.
 Here, $a^\dag$ ($a$) is the creation (annihilation) operator for the resonator mode coupling with $g_j$ to the $j$-th spin, which is described by the standard Pauli operators $\sigma_j^z$, $\sigma_j^+$, $\sigma_j^-$. Note that \eqref{eq:Hamiltonian} does not include direct dipole-dipole interactions, which -- although present in the actual sample -- do not seem to play a fundamental role for the formation of the echo pulses in our model. For large spin ensembles, we can use a mean-field formulation in the form of the Maxwell-Bloch equations for the resonator and spin expectation values \cite{Zens2019, SupplementaryInformation},
\begin{align} 
\label{eq:MaxwellBlocha}
&\frac{d}{dt}\langle a\rangle =-[\kappa_c+i(\omega_c-\omega_p)]\langle a\rangle-i\!\sum_{j=1}^Ng_j\langle\sigma_j^-\rangle\!+\eta(t), \\[-2mm] 
&\frac{d}{dt}\langle \sigma_j^-\rangle =-[\gamma_\perp+i(\omega_j-\omega_p)]\langle\sigma_j^-\rangle+i\,g_j\langle\sigma_j^z\rangle\langle a\rangle,\label{eq:MaxwellBlochb}\\[0.5mm]
&\frac{d}{dt}\langle\sigma_j^z\rangle =-\gamma_\parallel(\langle\sigma_j^z\rangle+1)
+2i\,g_j(\langle\sigma_j^-\rangle\langle a^\dag\rangle-\textit{c.c.}). 
\label{eq:MaxwellBlochc}
\end{align} 
Here, $\gamma_\perp=1/T_2$ ($\gamma_\parallel=1/T_1$) is the transverse (longitudinal) spin relaxation rate. We account for the dephasing of the spin ensemble by introducing the phenomenological Lorentzian spin spectral density, $\rho(\omega)=\{\pi\gamma_s[1+(\omega-\omega_s)^2/\gamma_s^2]\}^{-1}$, with width $\gamma_s$ and mean frequency $\omega_s$, characterizing the frequency distribution of the spin ensemble \cite{Angerer2017, Krimer2019,SupplementaryInformation}. For simplicity, we assume for the calculations presented in Fig.\,\ref{Fig2}\,(e) that all spins couple with the mean coupling strength $g_j=g_0=g_\mathrm{eff}/\sqrt{N}$ (in \cite{SupplementaryInformation} we discuss the impact of a distribution of $g_j$).  

To calculate the dynamics of the spin-resonator system, we numerically solve the Maxwell-Bloch equations \eqref{eq:MaxwellBlocha}-\eqref{eq:MaxwellBlochc} for two rectangular driving pulses with a width of $1\mu$s and $2\mu$s, a pulse delay of $\tau=80\mu$s, and a pulse amplitude of $\eta/\kappa_c=1.08\times10^5$. Furthermore, we set $\omega_c=\omega_p$, while the mean frequency of the spin ensemble is slightly detuned from the resonator frequency by $(\omega_s-\omega_c)/2\pi\!=\!\SI{0.14}{MHz}$ to match the experimental conditions in the strong-coupling regime (at $B_0\!=\!170.18$ mT). 

 The calculated average resonator photon number $|\langle a(t)\rangle|^2/N$ following an ordinary Hahn-echo sequence is presented in the upper panel of Fig.~\ref{Fig2}\,(e). Most importantly, we find that these numerical results nicely reproduce the multiple echo signatures found experimentally (see Fig.~\ref{Fig2}\,(d)), using only minimalistic assumptions. Additionally, these simulations provide the average spin expectation values $S_\text{av}^{x,y,z}:=\sum_j\langle\sigma_j^{x,y,z}\rangle/N$, which are not directly accessible in the experiment. From these quantities we can directly evaluate the macroscopic dipole moment $\sum_j\langle\sigma_j^{-}\rangle=N(S_{\mathrm{av}}^x+iS_\mathrm{av}^y)$, which couples the spin dynamics to the resonator field via \eqref{eq:MaxwellBlocha}. Hereby, we can directly confirm, e.g., that the arrival of the first conventional Hahn echo, at $t=0$, is accompanied by peaks in the average dipole moments $S_\text{av}^x$ and $S_\text{av}^y$, leading to a resonator-enhanced decay of the spin excitation $S_\text{av}^{z}$ (see also gray inset of Fig.\,\ref{Fig2}\,(e)). Confirming our heuristic model from above, the  same coincidence between peaks in the dipole moments of $S^x_{\rm av}$, $S^y_{\rm av}$, the steps in the decay of $S^z_{\rm av}$, and the emission of a photon pulse into the resonator is observed for all subsequent (unconventional) Hahn echos. This reduced model thus already reproduces all salient features of the experiment. As shown explicitly in \cite{SupplementaryInformation} the spin rotations on the Bloch sphere occurring during the emission of a Hahn echo are essential to produce the subsequent echo, a feature which is connected to strong spin-resonator coupling. We also checked in \cite{SupplementaryInformation} that imperfections in the second applied ($\pi$) pulse are required for the observation of multiple echos. Next steps in the improvement of the model shall include the dipole-dipole interactions between the spins, as well as the inclusion of the exact shape of the spectral spin and spatial coupling distributions.

In conclusion, we compared continuous-wave and pulsed ESR measurements on a weakly and strongly coupled spin ensemble using superconducting lumped element resonators. We observed a self-sustained train of periodic echo signatures after applying a Hahn echo sequence to the spin ensemble in the strong coupling regime and explain this effect using a simple model based on the inhomogeneous Tavis-Cummings Hamiltonian. Our work establishes a robust and self-sustained dynamical phenomenon in strongly coupled hybrid spin-photon systems, which may be relevant for quantum memory protocols.

\begin{acknowledgments} 
    We acknowledge stimulating discussions with K.~Lips, S.\,T.\,B.~Goennenwein,
    D.~Einzel and M.~Weiler and financial support from the Deutsche
    Forschungsgesellschaft via Germany's Excellence Strategy EXC-2111-390814868 and SPP 1601~(HU~1896/2\mbox{-}1). M.\,Z.\ and S.\,R.\ acknowledge support by the European Commission under Project NHQWAVE No. MSCA-RISE 691209 and by the Austrian Science Fund (FWF) through the Doctoral Program CoQuS (W1210). M.\,Z.\ would like to thank the Institute for Theoretical Atomic, Molecular, and Optical Physics (ITAMP) at Harvard for hospitality.
\end{acknowledgments}

\emph{Note added.---}During the preparation of the revised version of this manuscript, we became aware of a related work by Debnath \textit{et\,al.\,}\cite{Debnath:2020}.
%

\widetext
\clearpage
\begin{center}
\section{Supplementary Information}
\end{center}
\setcounter{equation}{0}
\setcounter{figure}{0}
\setcounter{table}{0}
\setcounter{page}{1}
\makeatletter
\renewcommand{\theequation}{S\arabic{equation}}
\renewcommand{\thefigure}{S\arabic{figure}}
\renewcommand{\bibnumfmt}[1]{[S#1]}
\renewcommand{\citenumfont}[1]{S#1}

\section{Effect of pulse imperfections and coupling strength}

Fully avoiding pulse errors in an inhomogeneously broadened spin ensemble is a challenging task that depends on the distribution of coupling strengths and spin frequencies as well as on the experimental circumstances. In particular, pulses of finite lengths and simple shapes typically result in imperfect rotation angles of individual spins in the experiment. However, as outlined in the main text, rotation errors are also proving to be an important part of the multi-echo formation process.

To determine the impact of rotation errors we present here additional simulations, where the action of the two microwave pulses in the Hahn echo sequence is included in the initial conditions of our theoretical model. The purpose of this procedure is to disentangle the intricate strong coupling dynamics during the $\pi/2$- and $\pi$-pulses from the subsequent dynamics. To be specific, we solve the Maxwell-Bloch equations for the initial conditions $\langle a \rangle=0$, $\langle \sigma_j^x \rangle=-\cos(\Delta_j\tau)\cos(\alpha)$, $\langle \sigma_j^y \rangle=-\sin(\Delta_j\tau)$, and $\langle \sigma_j^z \rangle=\cos(\Delta_j\tau)\sin(\alpha)$ at $t=0$, where $\Delta_j=\omega_j-\omega_p$ is the detuning between the spin frequency and the reference rotating frame, $\tau=20\,\mu$s is now the inter pulse delay, and $\alpha$ is the rotation angle of the second pulse. Note that these specific initial conditions correspond to a situation where all spins are collectively brought into the $xy$-plane using a perfect $\pi/2$-rotation along the $y$-axis for the first pulse. Then, after a free evolution time $\tau$, the spin ensemble is artificially rotated by an angle $\alpha$ along the $y$-axis. Note that with this procedure we effectively switch off the collective coupling between the spin ensemble and the resonator during this entire preparation period. As a result, we can study the impact of the spin-resonator coupling and rotation errors independently of the imperfections imposed by the microwave pulses.

\begin{figure}
\includegraphics[width=8.6cm]{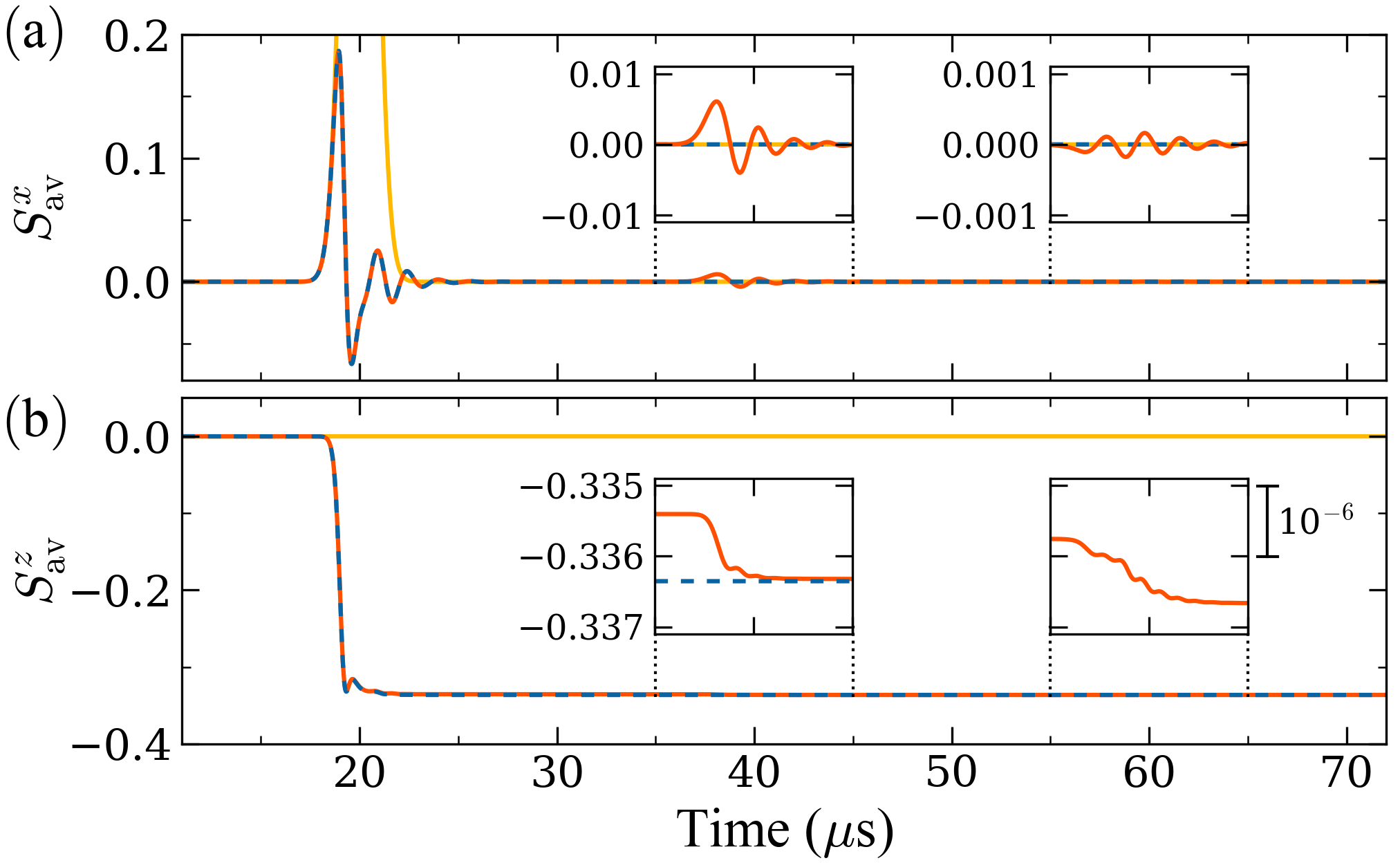}
\caption{Average spin expectation values $S_\mathrm{av}^{x,z}=\sum_j\langle \sigma_j^{x,z}\rangle/N$ versus time for a spin ensemble starting from an initial condition that imitates a Hahn echo sequence of a perfect $\pi/2$-rotation followed by an $\alpha$-rotation right before $t=0$. (a) $S_\mathrm{av}^{x}$ for a strongly coupled spin ensemble, $\Omega/2\pi=\SI{1.56}{MHz}$, after a perfect rotation $\alpha=\pi$ (blue dashed) and an imperfect rotation $\alpha=0.95\times\pi$ (red).  The imperfect rotation $\alpha=0.95 \pi$ is also shown for weak coupling $\Omega/2\pi=\SI{1.56}{ kHz}$ (yellow). The conventional Hahn echo at $t=\SI{20}{\mu s}$ is present in all situations, while additional echos at $t=\SI{40}{\mu s}$ and $t=\SI{60}{\mu s}$ (insets) are visible only for the combination of imperfect rotations and strong coupling. (b) Due to the strong coupling (blue and red) $S_\mathrm{av}^{z}$ changes significantly during the conventional Hahn echo. This effective rotation of the spin ensemble is absent for weak coupling (yellow). Much smaller but similar rotations are visible at $t=\SI{40}{\mu s}$ and $t=\SI{60}{\mu s}$ (insets) for    $\alpha=0.95 \pi$ (red), but not for $\alpha=\pi$ (blue). (Also in the right inset, the blue dashed line shows no variation, but falls outside of the zoom window.) }
\label{fig:pulse_errors}
\end{figure}

In Fig.~\ref{fig:pulse_errors} we present the results for the average spin expectation values $S_\mathrm{av}^{x,z}=\sum_j\langle \sigma_j^{x,z}\rangle/N$ for a Gaussian spin distribution, where we distinguish two different settings: (i) We consider a strongly coupled spin ensemble ($\geff/2\pi= 1.56\,\mathrm{MHz}$) and compare the evolution involving a perfect ($\alpha=\pi$) and a slightly imperfect ($\alpha=0.95 \pi$) refocusing pulse. (ii) We compare the dynamic evolution involving an imperfect rotation ($\alpha=0.95 \pi$) for strong ($\geff/2\pi= 1.56\,\mathrm{MHz}$) and very weak ($\geff/2\pi= 1.56\,\mathrm{kHz}$) coupling to the resonator. We first note, that the conventional Hahn echo at $t=20\,\mu$s is observed in $S_\mathrm{av}^{x}$, regardless of both the coupling strength and the rotation error. Next, we compare  the impact of the pulse rotation angle under strong coupling. While the results for the perfect and the imperfect rotation almost overlap during the conventional Hahn echo, additional echos at $t=40\,\mu$s and $t=60\,\mu$s arise only for $\alpha=0.95 \pi$, indicating that the pulse imperfections are relevant for the multi-echo formation. Staying with $\alpha=0.95 \pi$, but reducing the coupling strength to $\geff/2\pi= 1.56\,\mathrm{kHz}$ reveals the key role of the spin-resonator coupling. Although the spins build up a large dipole moment $S_\mathrm{av}^{x}$ during the conventional Hahn echo, the coupling to the resonator is too weak to cause a significant rotation of the spins on the Bloch sphere and therefore no visible echos are produced at later times. Our findings thus suggest that the enhanced rotation of the spins during the echos in combination with an imperfect refocusing pulse are the key building blocks for the formation of multiple echos.

\section{Variation of the pulse delay time $\tau$}

\begin{figure}
\includegraphics[width=8.6cm]{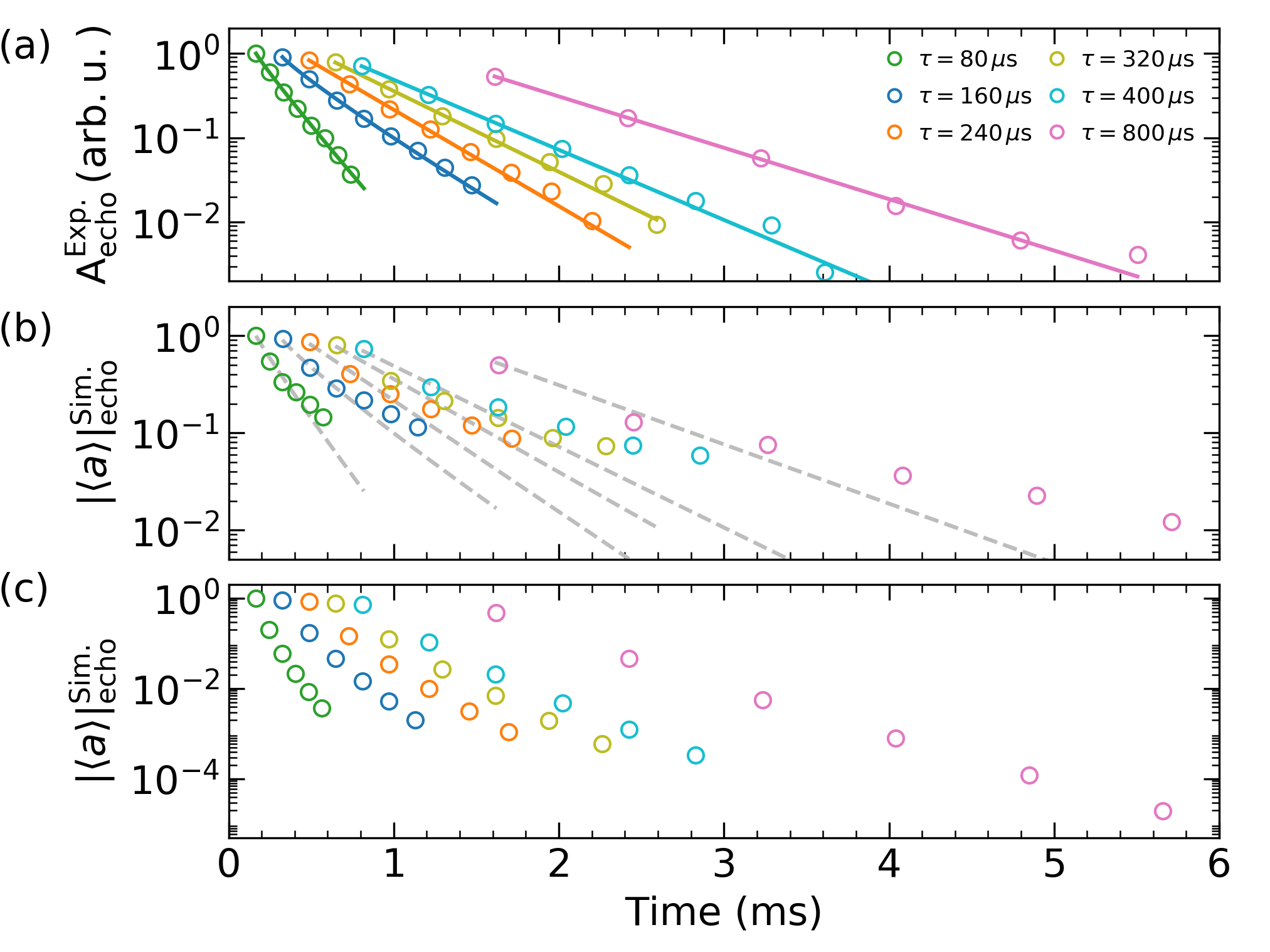}
\caption{(a) Experimental data: Integrated echo area for the individual echos as a function of time, shown for several echo spacings $\tau$. Solid lines are fits to an exponential decay law. (b,c) Simulation: Maxima of the individual echos as a function of time, shown for several echo spacings $\tau$ assuming (b) a Lorentzian and (c) a Gaussian spin distribution.  For comparison we show the fits to the experimental data again in (b) (gray dashed lines). All data points are normalized to the area/hight of the first echo of the $\tau=80\,\mu$s dataset.}
\label{fig:tau_var}
\end{figure}

One key parameter in the Hahn echo sequence is the inter-pulse delay $\tau$, which is varied in experiments to determine the coherence time of the spin ensemble. In particular, the analysis of the decay of the conventional Hahn echo gives access to this characteristic time. In this spirit, we present in Fig.\,\ref{fig:tau_var} the experimentally determined echo areas as a function of their arrival time for various $\tau$ recorded at a fixed magnetic feld of $170.18\,\mathrm{mT}$ using a wait time of $180\,\mathrm{s}$ between measurements. We find for the experimental data that the subsequent echos show a decreasing amplitude, which can be well described by an exponential decay (lines in Fig.\,\ref{fig:tau_var} (a)). The corresponding characteristic decay times $T_\mathrm{decay}$ increase for longer inter-pulse delays $\tau$. This can be rationalized by the observation that the formation of an echo constitutes an effective decay channel. Thus, we expect that $T_\mathrm{decay}$ should be fundamentally limited by the coherence time $T_2$, which is the case for the data presented here. In addition, we can compare the experimental observations with our theoretical model. In particular, we choose a Lorentzian and a Gaussian spin distribution of the same width $\gamma_s$ to study their impact on the echo decay. For both spin distributions the amplitude of the driving is chosen such that the first pulse corresponds to an effective $\pi/2$-rotation. On a first glance, we find that both spin distributions corroborate the experimental data, as both predict an initial exponential decay. However, we also find characteristic differences in the decay. For instance,  while the Gaussian-shaped distribution can be well described by an exponential decay, the Lorentzian-shaped distribution initially falls off with a fast rate and decays at later times with a noticeably smaller rate. On a quantitative level, the initial decay rates observed in the experiment are in reasonable agreement with the initial decay rates of the Lorentzian-shaped spin distribution (maximum deviation of $20\%$). For the Gaussian spin distribution the decay times of the individual echo trains exceed those observed in the experiment by approximately a factor of 10. In general, we note that the decay of the echo train does not only depend on $\tau$ and the characteristic parameters of the system, such as $\kappa$, $\kappa_\mathrm{ext}$, $\gamma_\mathrm{s}$, but also strongly depends on the exact shape of the spin distribution. In addition, we suspect that the dipole-dipole interaction present within the spin ensemble could additionally affect the characteristic decay time and speculate that the details of the experiment such as the spatial distribution of the excitation field $B_1$ as well as the amplitude and temporal shape the microwave pulses have the potential to modify this decay. 
A detailed analysis of these dependencies will be the subject of future work.

\section{Experimental Setup}
Below, we describe in detail the experimental setup used to obtain the results presented in the main text. We first describe the sample preparation, followed by a description of the cryogenic and room-temperature microwave circuitry. The sub-section \ref{demod} describes the digital down-conversion of the signal after digitization. Finally, we describe how we determine the integration window for the echo area integration of the echo trains.

\subsection{Sample preparation}
The sample investigated in the main part consists of two parts: a superconducting planar microwave resonator and a paramagnetic electron spin ensemble.

The microwave resonator is fabricated on top of a $6\times\SI{10}{mm^2}$ high-resistivity ($>\SI{10}{k\Omega cm}$) silicon substrate with natural isotope composition. The substrate is first cleaned in an ultrasonic bath using acetone and isopropyl alcohol. Then, a \SI{150}{nm} thick niobium layer is deposited onto the substrate in a sputter process. Next, the chip is spin-coated with photo resist and the resonator structure is defined via optical lithography. After development, the structure is transferred into the superconducting film using a reactive ion etching process. The chip is then placed into a gold-plated (oxygen-free highly-conductive) copper box and connected to this enclosure using conductive silver-glue at its boundaries. This forms the ground connection of the resonator. SMA end launch connectors are then inserted from both ends and the center pin of the end-launch is connected to the coplanar waveguide using silver glue.

As paramagnetic electron spin ensemble, we use phosphorus donors with a doping concentration of $\left[P\right] = \SI{1e17}{cm^{-3}}$ embedded in an isotopically purified \textsuperscript{28}Si host crystal with a residual \textsuperscript{29}Si concentration of \SI{0.1}{\%}. The \textsuperscript{28}Si:P crystal has a thickness of \SI{20}{\micro\meter} and was originally grown on top of a heavily boron-doped \textsuperscript{nat}Si substrate. An additional \SI{500}{nm} thick arsenic doped \textsuperscript{nat}Si layer was grown on top of the \textsuperscript{28}Si:P layer. We remove these additional layers by a combination of mechanical polishing and reactive ion etching. The resulting \SI{20}{\micro\meter} thick flakes are then placed with the utmost care on top of the resonator. The flakes are pressed onto the resonator using an additional piece of an \textsuperscript{nat}Si wafer and a PTFE screw in the lid of the sample box.

\subsection{Microwave circuit} 
\begin{figure}
    \includegraphics[width=3.0in]{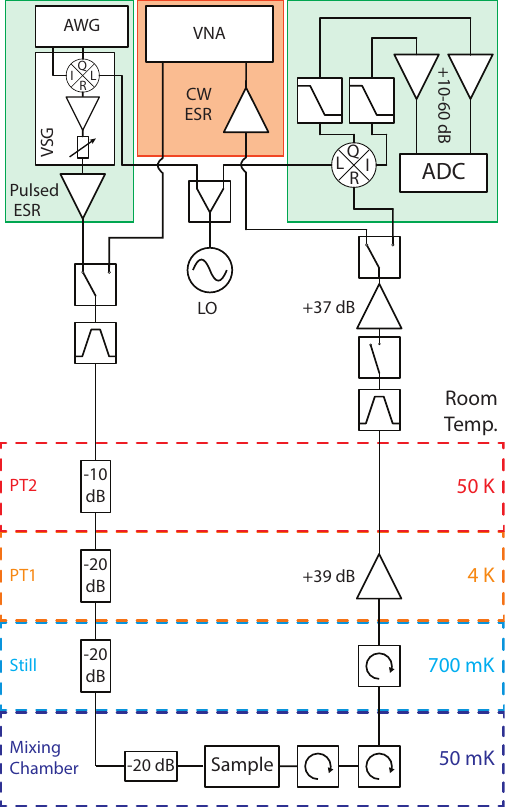}
    \caption{\label{MicrowaveSetup}Microwave setup for continuous-wave (red) and
    pulsed (green) electron spin resonance experiments. Details of the pulse
    bridge and detection scheme are given in the text.} 
\end{figure}

The microwave circuitry used in this work is presented in
Figure~\ref{MicrowaveSetup}.  

The main goal of the cryogenic microwave circuitry is to suppress room
temperature noise photons from reaching the sample under investigation. To this
end, the input lines are attenuated by \SI{70}{dB} at the various temperature
stages. On the output side, we use two cryogenic circulators on the mixing
chamber stage as well as one at the still level. The outgoing signal is
amplified by a cryogenic HEMT amplifier (Low Noise Factory LNC4\_8A) at the 4K
stage.

The microwave circuitry at room temperature to perform both continuous-wave (CW)
as well as pulsed ESR measurement via two latching electromechanical RF switches
(Keysight 8765B). The signal entering the cryostat is bandpass-filtered
(MiniCircuits VBFZ-5500-S+) to the relevant frequency range to reduce the power
load on the subsequent cryogenic stages. The output signal is bandpass-filtered
as well before entering a fast PIN diode switch (Analog Devices HMC-C019). This
switch blanks out the high-power microwave pulses from entering the sensitive
down-conversion setup. The signal is then further amplified at room-temperature
(B\&Z Technology BZP110UC1).

To perform CW ESR measurements, we connect a vector network analyzer (Rhode \&
Schwarz ZVA8) to the input and output line and measure the transmission
scattering parameter $\left|S_{21}\right|^2$.

Pulsed ESR measurements are performed using a in-house built microwave bridge.
We generate in-phase and quadrature signals of Gaussian-shaped pulses using a
fast arbitrary waveform generator (Agilent M8190A, \SI{12}{GS/s}) at an
intermediate frequency of $f_\mathrm{IF} = \SI{42.5}{MHz}$. The pulses are then
up-converted to the resonance frequency using a vector signal generator (Rhode \&
Schwarz SGS100A) and further amplified (CTT AGX0218-3964) before reaching the
input side of the cryostat. The pulse power  before the microwave switch at the input of the cryostat is
$+\SI{25}{dBm}$, resulting in a maximum echo signal for pulse times of
\SI{1}{\micro\second} and \SI{2}{\micro\second}.

Detection of the resulting spin echos is performed by a heterodyne
down-conversion setup. The signal is down-converted using an IQ mixer (Marki
IQ-0307L). The down-converted signal with frequency $f_\mathrm{IF}$ is then
lowpass-filtered to reduce LO leakage. The signal is  amplified with variable
gain between $10$ and $\SI{60}{dB}$ (FEMTO DHPVA-200) to utilize the full dynamic
range of the analog-to-digital converter (Spectrum M4i.4451-x8). The digitizer
card records both the in-phase and quadrature component at a sample rate of
\SI{500}{MS/s}.

To ensure a stable phase synchronization between the devices, all devices are
synchronized using an oven-stabilized 10 MHz reference signal (Stanford Research
Systems FS725). The LO signal (Agilent E8257D) is provided to both the vector
network source as well as the IQ mixer using a power divider.

\subsection{Signal demodulation}\label{demod}
In this section we describe our algorithm to demodulate the signal at the
intermediate frequency (here $f_\mathrm{IF} = \SI{42.5}{MHz}$) to baseband (DC).
We first calculate the complex signal $Z = I(t) + iQ(t)$ from the recorded
in-phase and quadrature signal. The microwave transmission signal $S_{21}$ is obtained by multiplying $Z$ with a complex sinusoidal
\begin{equation} 
    S_{21} = Z \cdot \exp\left(-i\left(2\pi f_\mathrm{IF} +
\phi\right)\right), 
\end{equation} 
where $f_\mathrm{IF}$ is the intermediate
frequency and $\phi$ is the demodulation phase. This shifts the frequency of the
signal to the baseband. We choose $\phi$ in such a way that the signal in the
real part of $S_{21}$ is maximized. After the frequency conversion, we apply a
lowpass filter (digital Butterworth filter of 5th order) with a cutoff frequency
of \SI{10}{MHz} and re-sample the signal at a sample rate of \SI{20}{MS/s} to
reduce the file size of the measured signals.

\subsection{Echo integration}
In the following we describe our procedure to integrate the echo signal. The key here is to determine the length of the integration window, $\Delta t$, given
the following two challenges:
\begin{enumerate} 
    \item For short $\tau$, we cannot use a very broad integration window, as
        the echo peaks are close to each other. Therefore, the integration
        window has to be chosen for each value of $\tau$ individually.
    \item As we integrate the magnitude of the signal $|S_{21}|$, there is a finite
        DC offset $V_\mathrm{offset}$ present in the signal. This offset adds a
        finite contribution $V_\mathrm{DC}\cdot\Delta t$ to the integrated echo
        area, where $\Delta t$ is the length of the integration window.  
\end{enumerate} 
Our algorithm works as follows: First, we determine all echo peaks using a
peak-detection algorithm. In the next step, we integrate the signal using a
numerical trapezoid integration, centered around the second detected echo peak
with varying integration window $\Delta t$. When plotting $A_\mathrm{echo}$ as a
function of $\Delta t$, we can distinguish three regions (c.f.\
Figure~\ref{IntegrationWindow}): 

\begin{figure}
    \includegraphics{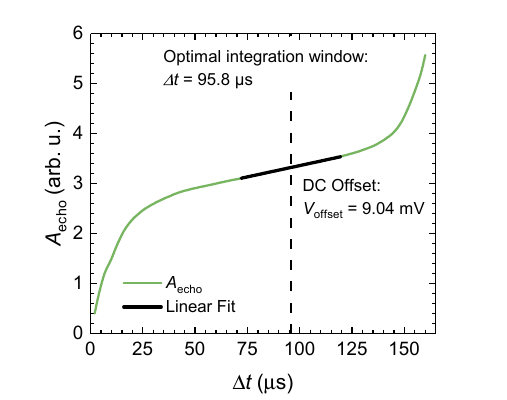}
    \caption{\label{IntegrationWindow}Determining the integration window $\Delta
t$ and DC offset $V_\mathrm{offset}$. For details see text.} 
\end{figure} 

A steep increase for small (large) values of $\Delta t$. These are caused by
partial integration of the investigated (next) echo peak. In the intermediate
region, we observe a linear increase of $A_\mathrm{echo}$ with $\Delta t$. Here,
the investigated echo peak is completely inside the integration window and the
increase of the echo area is due to the integration of the DC offset. To
determine the optimal integration window, we calculate the minimum of the first
derivative $\mathrm{d}A_\mathrm{echo}/\mathrm{d}\Delta t$ (dashed line in
Figure~\ref{IntegrationWindow}). The DC offset is determined as the slope of a
linear fit in the linear regime (solid line in Figure~\ref{IntegrationWindow}).

For the final integration, we subtract the DC offset from the magnitude signal
and integrate each detected echo peak using the previously determined
integration window.

\section{Spin-resonator coupling}
In this section, we discuss the spin-resonator coupling. The planar resonator
structure used in our experiments creates an inhomogenous microwave magnetic field and
leads therefore to a distribution of the spin-resonator coupling rate. The
analysis of the coupling rate in the presence of an inhomogeneous microwave magnetic field
distribution is based on our work described in Ref.\,\cite{Weichselbaumer2019}.

\begin{figure*}
    \includegraphics{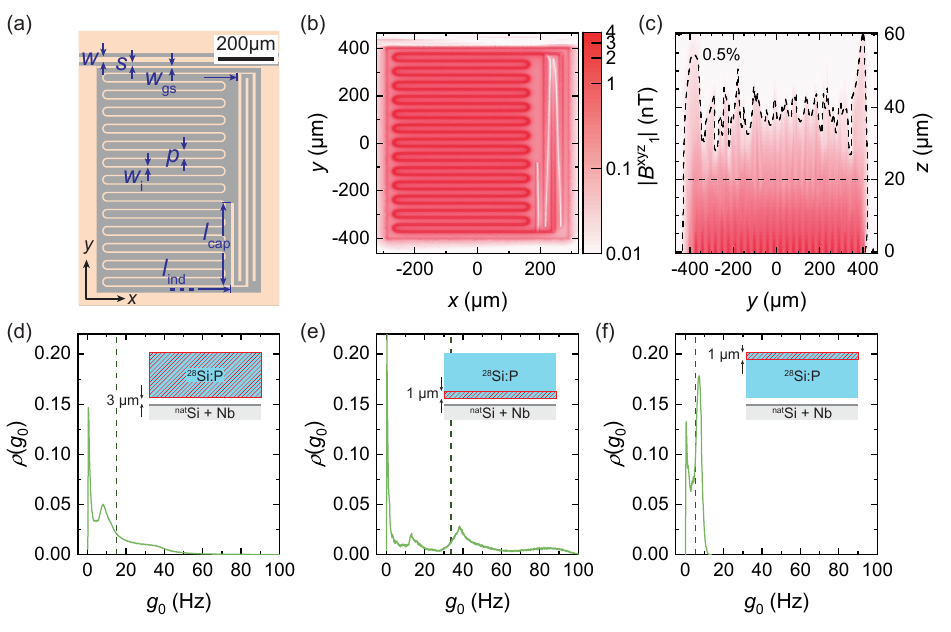}
    \caption{\label{ResonatorAnalysis} Analysis of the resonator used in the
        experiment. (a) Schematic of the lumped element resonator. (b) Field
    distribution in a top-view. (c) Field distribution in the $yz$-plane for $x
    = 0$. The dashed line indicates the region where the amplitude decayed to
    \SI{0.5}{\%} of the maximum amplitude. (d)-(f) Distribution of the
    collective coupling rate for (d) $0 < z \leq \SI{20}{\mu m}$, (e) $0 < z \leq\SI{1}{\mu m}$
     and (f) $\SI{19}{\mu m} < z \leq \SI{20}{\mu m}$.
    The dashed line indicates the average coupling strength in this sample
    region.}
\end{figure*}

A schematic of the resonator used in the experiment is displayed
Fig.~\ref{ResonatorAnalysis}~(a). The resonator is embedded in the ground plane
of a coplanar waveguide (signal line width $w = \SI{20}{\mu m}$, gap width $s =
\SI{12}{\mu m}$). The resonator is separated from the signal line by a screening
line (width $w_\mathrm{gs} = \SI{10}{\mu m}$), which defines the external
coupling rate\,\cite{Weichselbaumer2019}. The resonator consists of an inductor
(wire width $w_\mathrm{i} = \SI{5}{\mu m}$, pitch distance $p = \SI{20}{\mu m}$)
with a total length of $l_\mathrm{ind} = \SI{11.35}{mm}$ and a finger capacitor.
By changing the length $l_\mathrm{cap}$ of the capacitor finger the resonance
frequency can be tuned.

For a further analysis, we perform finite element simulations using CST
Microwave Studio 2016\,\cite{CST2016} to extract the three-dimensional microwave magnetic field
distributions of the resonator.  Figures\,\ref{ResonatorAnalysis}\,(b) and (c) show
the spatial distribution of the magnitude of the vacuum magnetic field
fluctuations $|B_1^{yz}|$, i.e.\ the field component that is perpendicular to
the static magnetic field $B_0$ along the $x$-direction. The data is exported from
CST Microwave Studio in volume elements of $1\times 1\times\SI{1}{\mu m^3}$. The
dashed line in panel~(c) marks the region where the field amplitude decayed to
\SI{0.5}{\%} of the maximum field amplitude.
We define this volume as the mode volume of the resonator $V_\mathrm{m} =
\SI{1.41e-11}{m^3}$.  Due to the anti-parallel current flow in the inductor
wires, the dynamic magnetic field interferes destructively in the far-field.
This limits how far the magnetic field reaches into the $z$-direction and
enhances the sensitivity of the resonator to spins close to the superconducting resonator.

The single spin-resonator coupling is given by\,\cite{Wesenberg2009} 
\begin{equation}
    g_0 = g_\mathrm{s}\mu_\mathrm{B}B_{1,0} /\hbar,
    \label{eq:single-spin-coupling}
\end{equation}
where $g_\mathrm{s} = 1.9985$ is the
electron g-factor of phosphorus donors in silicon\,\cite{Feher1959} and $\mu_\mathrm{B}$ is the Bohr magneton.
$B_{1,0}$ describes the magnetic field generated by vacuum fluctuations in the
resonator. $B_{1,0}$ is given by\,\cite{Schoelkopf2008} $B_{1,0} =
\sqrt{\mu_0\hbar\omega_r/(2V_\mathrm{m})}$, where $\mu_0$ is the vacuum
permeability, $\hbar$ is the reduced Planck constant and $\omega_r$ is the
resonance frequency of the resonator. Collective coupling effects lead to an
enhancement of the single-spin coupling rate by a factor $\sqrt{N}$, where $N$
is the number of spins. Thus, the collective coupling strength is given as
\begin{equation}
    g_\mathrm{eff} = \frac{g_\mathrm{s}\mu_\mathrm{B}}{2\hbar}
                     \sqrt{\frac{1}{2}\mu_0\hbar\omega_\mathrm{r}\rho_\mathrm{eff}\nu}.
     \label{eq:geff}
\end{equation}
In this expression, the number of spins, $N$, is replaced by $N =
\rho_\mathrm{eff}V = \rho P(T)V$, where $\rho$ is the donor concentration, $P(T)$
is the thermal spin polarization and $V$ is the sample volume. The filling
factor $\nu = V/V_\mathrm{m}$ describes the ratio between the sample volume and
the mode volume of the resonator. 

The planar resonator structures used in this experiment generate an
inhomogeneous microwave magnetic field $B_1$, which has to be taken into account
in the filling factor 
\begin{equation}
    \nu = \frac{\int_\mathrm{Sample} B_1^2(\vec{r})\,\mathrm{dV}}
               {\int_\mathrm{Mode} B_1^2(\vec{r})\,\mathrm{dV}}.
\end{equation}
We can calculate the filling factor from the exported three-dimensional
distribution of the microwave magnetic field using the expression
\begin{equation}
    \nu = \frac{\sum_V \left|B_\mathrm{1,sim}^{yz}(\vec{r})\right|^2}
    {\sum_{V_\mathrm{m}} \left|B_\mathrm{1,sim}^{xyz}(\vec{r})\right|^2}.
\end{equation}

With this approach, we obtain a theoretically expected spin-resonator coupling of
\SI{2.33}{MHz}, which somewhat over-estimates our experimentally defined value. We
explain this by a small gap between the resonator and the spin
sample\,\cite{Zollitsch2015}. Assuming a gap of \SI{2.91\pm0.02}{\micro\meter}, we obtain
a quantitative agreement between the theoretically expected spin-resonator coupling and the
experimentally determined value of \SI{1.54}{MHz}. However, we want to emphasize, that  one ingredient for the observation of the phenomenon of a self stimulated echo train is a sufficiently large coupling rate $\geff$, i.e.  placing the system in the strong coupling regime. 

Using Eq.~\ref{eq:single-spin-coupling} we can calculate the distribution of the
single spin-resonator coupling $g_0$. We present the data in
Fig.~\ref{ResonatorAnalysis}~(d) to (f) for different regions above the
resonator. Note that we included the finite gap between the resonator and the
sample in these calculations. Panel~(d) presents the coupling distribution over
the entire sample region with a mean coupling strength of $g_\mathrm{0,mean} =
\SI{14.93}{Hz}$ (dashed line). This results in a number of spins contributing to
the signal according to $N \approx (g_\mathrm{eff}/g_\mathrm{0,mean})^2 =
\num{1.06e10}$.  For a thin layer of the spin ensemble facing the resonator we compute an enhanced 
single spin-resonator coupling strength with a mean value of \SI{33.74}{Hz}, while spins on the
opposite side (panel (f)) couple relatively weakly with on average \SI{5.33}{Hz}. 
The low-frequency peak in the coupling distribution can be attributed to spins outside the
resonator dimensions, at the edges of the sample.

\section{Estimate of the driven Rabi frequencies and pulse lengths}

The finite element simulation of the microwave resonator also allows us to estimate the microwave $B_1$ fields present during the microwave pulses and correspondingly the expected pulse durations for the $\pi/2$ and $\pi$ pulses. 
In a  simplifying estimate, we can utilize the computed $g_0$ from Fig.\,\ref{ResonatorAnalysis} to estimate the driven Rabi frequency $\omega_1$, as the latter is given by $g_0 \sqrt{n_\mathrm{c}}$ (cf. Eq.\,\ref{eq:single-spin-coupling}) \cite{Chiorescu:2010hw, Weichselbaumer2019, Bienfait2016a}. For an initial estimate for $n_\mathrm{c}$, we turn to the Maxwell-Bloch equations and in particular (\ref{eq:MaxwellBlocha}). In detail, we relate the driving amplitude $\eta=\sqrt{\frac{2 \kappaext P_\mathrm{mw}}{\hbar \omega_\mathrm{c}}}$ to the experimental microwave power $P_\mathrm{mw}$. For a coarse estimate, we further assume a resonant excitation of the microwave resonator with the external microwave tone ($\Delta_\mathrm{c}=0$) and neglect the modifications of the microwave susceptibility of the system stemming from the strong coupling between the spin ensemble to microwave radiation. Note, that these reduce the photon number $n_\mathrm{c}$ in a complex fashion, and hence we expect to overestimate our driven Rabi frequency.  Using the parameters given in the main text, we find $n_\mathrm{c}=\SI{2.1e10}{}$ for a peak microwave power of +25dBm at the input of the dilution refrigerator, where we assume that attenuation is solely given by the microwave attenuators presented in Fig.\,\ref{MicrowaveSetup} (a total of 70dB attenuation).

In the driven Rabi regime, we next quantitatively estimate $\omega_1$ by $g_0 \sqrt{n_\mathrm{c}}$. Using the peak in Fig.\,\ref{ResonatorAnalysis}\,(d) at $g_0/2\pi=\SI{8}{Hz}$, we obtain $\omega_1/2\pi=\SI{1.2}{MHz}$ corresponding to a $\pi/2$-time of $\SI{200}{ns}$. This is a factor of 5 shorter than our experimentally chosen $\pi/2$ time, however it is worth to point out that this estimate is purely based on the design parameters of the resonator and the attenuators mounted in microwave delivery lines in the setup. Hence, this estimate neglects the additional input losses of the microwave lines, the insertion-loss of the microwave switch and the band pass filter as well as cable connectors, all of which are part of the microwave  input circuitry. Those will  further reduce the input power supplied to the resonator (we estimate this to be of the order of 5-10dB, corresponding to a reduction in $\omega_1$ between a factor of roughly 2-3). In addition, this estimate also neglects the modified transmission when the spin ensemble is set in resonance with the microwave resonator. In summary, our crude estimate for the pulse durations for a $\pi/2$ and $\pi$ pulse agrees well with our selected pulse times. Moreover, this estimate also emphasizes that the pulses have a significant $B_1$ distribution as can be seen in Fig.\,\ref{ResonatorAnalysis}\,d). 

\section{Experimental Pulse optimization}

Experimentally, we optimize the pulse angles via the detected echo amplitude. In detail, we vary the pulse length of the first pulse $t_\mathrm{duration}$ and second pulse $2\cdot t_\mathrm{duration}$ until we observe a maximum in the echo amplitude. Although this analysis does not give  direct information about the pulse angles of the first and second pulse, we experimentally notice that our pulse settings allow for a partial inversion of the echo, as seen in Sec.\,\ref{T1measurement}. This observation suggests that we indeed obtain a rotation angle of the order of $180^\circ$ for our effective $\pi$-pulse and hence confirms the rotation angles of the order of $90^\circ$ for our effective $\pi/2$-pulse.

\section{Phase cycling experiments}

\begin{figure*}
    \includegraphics[]{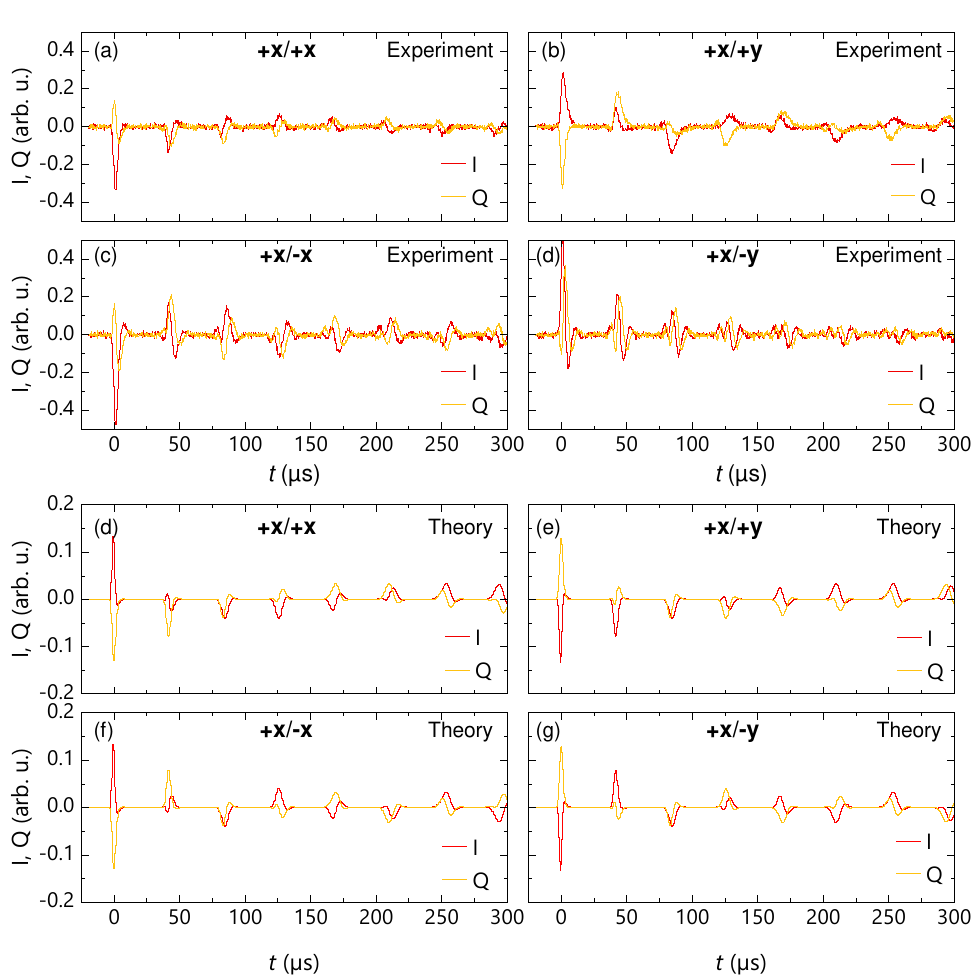}
    \caption{\label{phase-cycling}Phase cycling measurements. Quadratures $I$ and $Q$ of the recorded and simulated microwave transmission for (a), (e)~$+x/+x$, (b), (e)~$+x/+y$ (\ang{90} phase shift), (c), (f)~$+x/-x$ (\ang{180} phase shift) and (d), (g)~$+x/-y$ (\ang{270} phase shift). The magnetic field was set to the low-field hyperfine transition, which is strongly coupled to the microwave resonator. The echo signal is contained in both microwave signal quadratures and no clear phase relation between subsequent echos is visible.}
\end{figure*}

The experimental data in the main text were recorded with a ``$+x/+x$'' pulse sequence, i.e.\ the two microwave pulses are in phase. We have additionally recorded echo trains where a relative phase shift between the two pulses has been applied. In order to verify the occurence of the echo train phenomenon, a second sample has been used, which is nominally identical to the sample used in the main text. The experiments were performed with the magnetic field centered on the low-field hyperfine transition of the phosphorus donors, which is strongly coupled to the microwave resonator. In Fig.~\ref{phase-cycling}, we show the recorded quadratures, $I$ and $Q$ of the microwave transmission signal as a function of time for (a)~$+x/+x$, (b)~$+x/+y$ (\ang{90} phase shift), (c)~$+x/-x$ (\ang{180} phase shift and (d)~$+x/-y$ (\ang{270} phase shift). In contrast to conventional ESR experiments, where the ESR signal is typically contained in a single phase, in the strong coupling regime the microwave signal is contained in both quadratures. Additionally, no clear phase relation between subsequent echos is observable but rather a phase rotation from one echo to the next. We plot the quadratures of the simulated microwave transmission signal for $+x/+x$ and $+x/+y$ in panel (e) and (f), respectively. Our simulations can qualitatively reproduce the complicated phase relation of the echos.

\section{Conventional \texorpdfstring{T\textsubscript{2}}{T2} measurements } \label{T2measurement}
In a conventional ESR experiment, the coherence time $T_2$ is measured by Hahn
echo spectroscopy. A series of Hahn echo pulse sequences consisting of two
pulses are performed, where the pulse spacing $\tau$ is varied. The resulting
echo appearing $\tau$ after the refocussing pulse is digitized and integrated. The echo area $A_\mathrm{echo}$ then decreases with the characteristic coherence time $T_2$ in an exponential fashion. We use this experimental approach to determine the coherence time $T_2$.

\begin{figure*} 
    \includegraphics{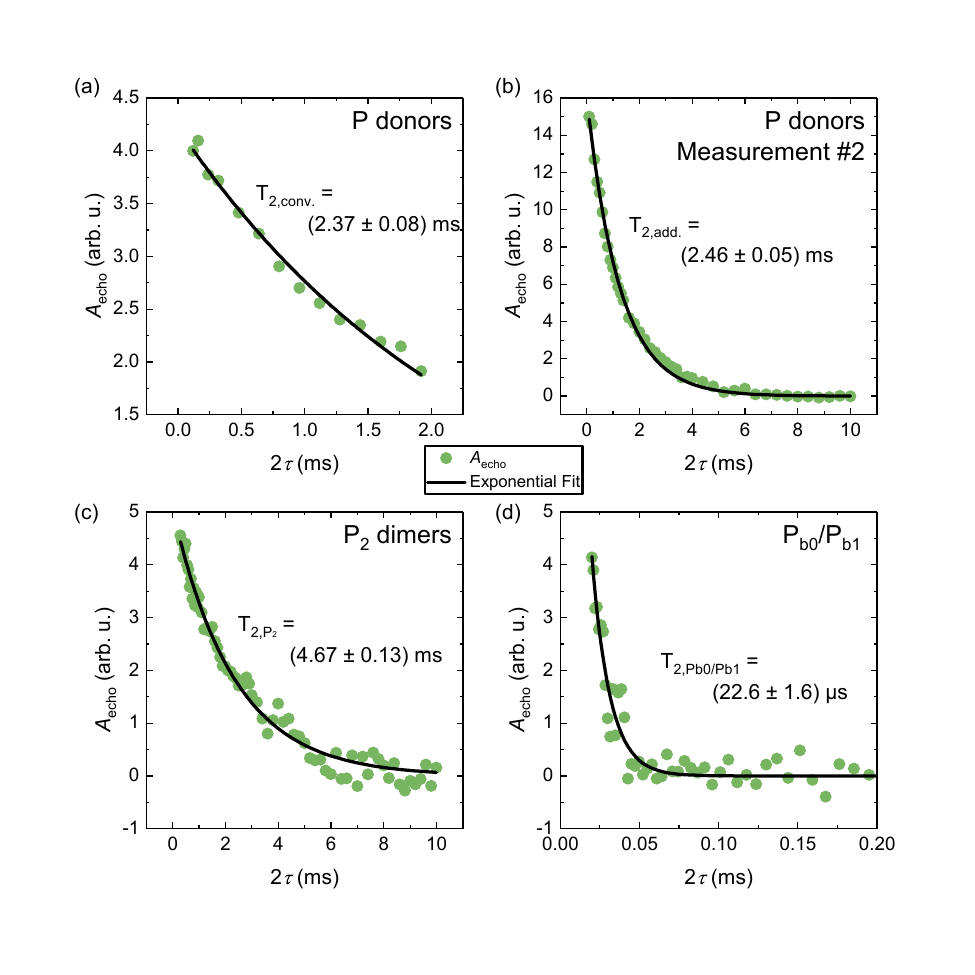}
    \caption{\label{T2Conventional}Determination of the coherence time $T_2$
        using conventional Hahn echo spectroscopy for the individual donors with
        \textbf{(a)} the same data as presented in the main text, \textbf{(b)}
        data from an additional measurement, where we varied $\tau$ and only
        recorded the first echo,  \textbf{(c)} the $P_2$ dimers, and (d) the $\mathrm{P_{b0}/P_{b1}}$ defects. For details
        see text.
    } 
\end{figure*}

In Figure~\ref{T2Conventional}, we show such conventional $T_2$ measurements of the
spin ensembles in our sample. Panel (a) shows the integrated echo area of the
first (conventional) echo of the data presented in Fig.\,\ref{fig:tau_var}\,(a). The exponential
fit (solid line) results in $T_{2,\mathrm{conv.}} = \SI{2.37\pm 0.08}{ms}$. As
this fit contains only a small number of points due to the limited $\tau$
resolution, we have performed an additional measurement for increased $\tau$,
where we have only digitized the first echo. The evaluation of the $T_2$ time
for this measurement presented in panel~(b) results in $T_{2,\mathrm{add.}} = \SI{2.46\pm
0.05}{ms}$, which is in agreement with the first measurement.  In panel~(c), we
present the same measurement as in (b), with the magnetic field set to the
resonance field of the $P_2$ dimer transition. Here, we extract a coherence time
$T_{2,P_2} = \SI{4.67\pm 0.13}{ms}$. Panel~(d) shows the coherence time
measurement of the P$_\mathrm{b0}$/P$_\mathrm{b1}$ defects with
$T_{2,\mathrm{P}_\mathrm{b0}/\mathrm{P}_\mathrm{b1}} = \SI{22.6\pm1.6}{\mu s}$.

\begin{figure} 
    \includegraphics{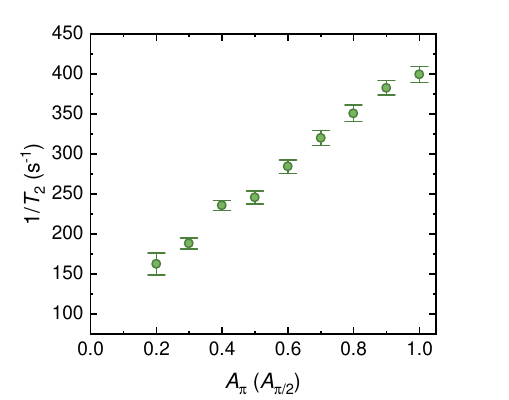}
    \caption{\label{T2InvDonorsAmplitudeSweep} $T_2$ measurement with variable
    amplitude of the second pulse. By decreasing the effective flipping angle in
the second pulse, instantaneous diffusion effects are reduced and $T_2$
increases. } 
\end{figure}

In samples with a large donor concentration as in our case, it is expected that the $T_2$ time is
limited by instantaneous diffusion, originating from a dipole-dipole interaction
between neighboring spins\,\cite{Klauder1962,Tyryshkin2003}. The influence of
instantaneous diffusion on the $T_2$ time can be reduced by reducing the
flipping angle of the second pulse in the Hahn
echo\,\cite{Tyryshkin2003,Shankar2015}. We performed $T_2$ measurements and
reduced the amplitude of the second pulse $A_\pi$ in relation to the amplitude
of the first pulse, $A_{\pi/2}$. We plot the inverse time $1/T_2$ in
Fig.~\ref{T2InvDonorsAmplitudeSweep}. We observe that $T_2$ increases when
decreasing the effective flipping angle showing a maximum $T_2$ of  $\SI{6.14}{ms}$. The trend, that smaller rotation angles have a positive effect on $T_2$ is compatible with the mechanism of instantaneous diffusion. Nevertheless, one would expect that the inverse $T_2$ time scales with  $\sin(\Theta/2)^2$, where $\Theta$ is the rotation angle of the second pulse \cite{Salikhov:1981ea, Tyryshkin:2011fi}. However, Fig.\,\ref{T2InvDonorsAmplitudeSweep} does not display this functional behavior, but rather a linear dependence on the pulse amplitude $A_\pi$. We speculate, that the details of the complex $B_1$ distribution and the spectral distribution of the spin ensemble $\rho(\omega)$ might be at the origin of this observed behavior.    

The coherence times reported here are exceptionally long  compared to conventional pulsed ESR experiments at higher temperatures \cite{Tyryshkin2003, Shankar2015}. We suspect that the long coherence times, which we find already for the initial Hahn-echo sequence, are a result of the suppression of instantaneous diffusion. As reported by Taylor et al. \cite{Taylor:1974}, long and weak amplitude pulses cause an effective increase of the $T_2$ time by selecting only a part of the ESR transition and hereby causing a suppression of instantaneous diffusion. In a reference experiment, we performed standard measurement of the coherence time with a Hahn-echo sequence at 6\,K (in a commercial Bruker ESR system) and find  $T_2\approx\SI{30}{\mu s}$, which is in good agreement with e.g. Ref.\,\citep{Tyryshkin2003}.

\section{\texorpdfstring{T\textsubscript{1}}{T1} measurements \label{T1measurement}} 
\begin{figure} \includegraphics{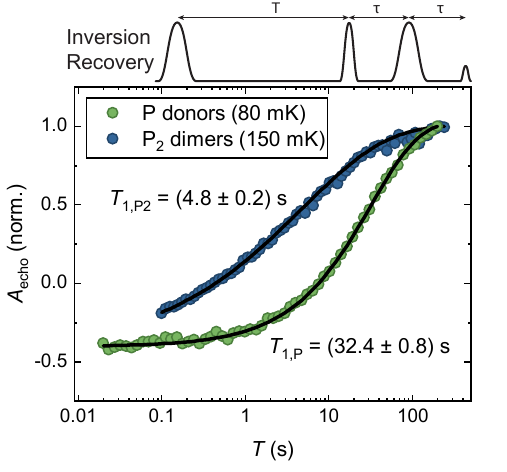}
    \caption{\label{T1Donors} $T_1$ measurement using an inversion recovery pulse
    sequence (top). Due to the non-ideal inversion the curve is not symmetric to
zero. The solid line is a fit to Eq.~(\ref{eq:t1exp}).} 
\end{figure}

To measure the spin life time $T_1$ we use an inversion recovery pulse sequence\,\cite{Schweiger2001}, as shown in the top of
Fig.~\ref{T1Donors}. Conceptually,  the first pulse in this three-pulse sequence inverts the spin ensemble. After a variable wait time $T$ a standard Hahn echo with fixed $\tau$ is used to probe the magnetization along the $z$-axis, giving a measure of the $T_1$ time.

In Fig.~\ref{T1Donors} we plot the extracted echo area as a function of the wait time $T$ for both the individual P donors and the P$_2$ dimers. Note that the measurements have been recorded at an elevated temperature compared to the measurements in the main text, which, however, has only a small impact on the determined value.

For small $T$, the partially inverted spin ensemble has a net moment along the $-z$ axis and
the resulting echo is negative. Ideally, the inversion pulse should result in a normalized echo amplitude of $-1$ for $T=0$, which is not the case here, probably due to the distribution of $B_1$ excitation fields. With increasing $T$ spins relax back to thermal
equilibrium along $-z$ and the echo area increases. We fit the following
function based on a stretched exponential to the data to extract the $T_1$ time: 
\begin{equation} 
A_\mathrm{echo}
= y_0 + A\cdot\left[1 - 2\exp\left(-\left(T/T_1\right)^b \right) \right].
\label{eq:t1exp} 
\end{equation} 
From this fit we extract $T_\text{1,P} =
\SI{32.4\pm0.8}{s}$ with $b=0.75$ for the low-field hyperfine split transition. For the P$_2$ dimers we extract $T_\text{1,P\textsubscript{2}} = \SI{4.8\pm0.2}{s}$ and $b=0.43$.
A stretched exponential form of the relaxation has been reported, e.g., in NMR for a superposition of single-exponential decays\,\cite{Alaimo1997}. As the Purcell-enhanced relaxation process depends on the spin-resonator coupling\,\cite{Bienfait2016a}, which is highly inhomogeneous in our case, we obtain a distribution of  relaxation times, justifying the use of a stretched exponential. Note that we introduce an additional offset $y_0$ in Eq.~\ref{eq:t1exp} to
account for non-ideal inversion due to the inhomogeneous $B_1$ field
distribution. From the ratio $y_0/(y_0 + A)\approx 0.338$ of the phosphorus donors, we can estimate that we effectively invert about \SI{34}{\%} of the addressed spin ensemble.

We next discuss two mechanisms which could account for these rather short relaxation times: (i) The shortening of the $T_1$ time due to Purcell enhancement and (ii) the one-phonon relaxation process.

\textit{Purcell-enhanced \texorpdfstring{T\textsubscript{1}}{T1} times ---} One mechanism resulting in an enhanced energy relaxation time is Purcell enhancement. This mechanism is present during the free evolution time of the experiment, where each spin individually couples to the microwave resonator. Bienfait et al.\,\cite{Bienfait2016a} discussed this as function of the detuning $\delta$ of the microwave resonator from the spin systems and find for bismuth donors in silicon shortened relaxation times in the seconds range. Following their discussion, we can calculate the Purcell rate by
\begin{equation}
    \Gamma_\mathrm{P}=(2\kappac) \frac{g_0^2}{(2 \kappac)^2/4 + \delta},
\end{equation}
where we have replaced the FWHM $\kappa$ of Ref.\cite{Bienfait2016a} with our HWHM $\kappac$. Using  the peak in $g_0/(2\pi)=\SI{8}{Hz}$ depicted in Fig.\,\ref{ResonatorAnalysis}\,(d) and $\kappac/(2 \pi)=\SI{565}{kHz}$ of the main text, we expect a Purcell-limited $T_1$ time of $\SI{700}{s}$. However, we also find a considerable amount of spins with a spin-resonator coupling of $\SI{40}{Hz}$, which would translate to a $T_1$ time of $\SI{30}{s}$. We speculate that spatial diffusion \cite{Bloembergen:1949gv,Eberhardt:2007fh,Redfield:1959hc,Vugmeister:1976co} can then assist with the relaxation of the majority of spins in the mode volume. However, tailored experiments, which are beyond the scope of this work, will be required to test this conjecture. 

\textit{One-phonon relaxation process --- } In addition, we can consider the $T_1$ process originating from the relaxation with the phonons. As our experiments are performed at low temperatures, we can reduce the discussion to the one-phonon relaxation process \cite{Feher1959b}. Morello et al. \cite{morello2010} discussed this process, which was initially presented by Hasegawa et al. \cite{Hasegawa:1960ey} in the low temperature limit. Both report for $g_s \mu_B B \ll k_\mathrm{B} T$ a temperature and magnetic field dependence of the spin lattice relaxation rate of 
\begin{equation}
\frac{1}{T_1}\propto  B^4 T.
\end{equation}
To discuss the phonon related relaxation process at even lower temperatures, we need to account for the phonon population, which is given by the Bose-factor $n_\mathrm{phonon}=1/(\exp(g_s \mu_\mathrm{B} B/k_\mathrm{B}T)-1)$ \cite{Hasegawa:1960ey}. Then 
\begin{equation}
\frac{1}{T_1} \propto B^5 n_\mathrm{phonon} \left(1+ \exp(g_s \mu_\mathrm{B} B/k_\mathrm{B}T)\right)
\end{equation}
For $g_s \mu_\mathrm{B} B \gg k_\mathrm{B} T$, this simplifies to the expected $B^5$ dependence, while for $g_s \mu_\mathrm{B} B \ll k_\mathrm{B} T$ we find the limit of $T_1^{-1}\propto B^4 T$. Using the reported spin relaxation time by Feher and Gere \cite{Feher1959b} for our donor concentration of $[P]=\SI{1e17}{cm^{-3}}$ of $T_1=\SI{1}{s}$ at $B=\SI{0.32}{T}$ as a calibration point, we can now extrapolate to our experimental temperature $T=\SI{100}{mK}$ and $B=\SI{0.17}{T}$. We find $T_1=\SI{110}{s}$. We note that this relatively short $T_1$ time is mostly caused by the high doping concentration.

In summary, both presented relaxation mechanisms reasonably explain our measured spin relaxation times $T_1$. 

In addition, we can use these estimates to calculate the expected spectral diffusion rate, which might mask the $T_1$ measurement and has potentially impact on the experimentally determined  $T_2$ times presented in this paper. Spectral diffusion depends on the donor concentration $[\mathrm{P}]$ and the corresponding time constant is given by \cite{Tyryshkin:2011fi}
\begin{equation}
    T_{\mathrm{SD}}=\sqrt{ \frac{18\sqrt{3}}{\mu_0} \frac{ \hbar}{(g_\mathrm{s} \mu_\mathrm{B})^2} \frac{T_1}{[P]}}
\end{equation}
Using the $T_1$ times determined above of $\SI{700}{s}$ and $\SI{110}{s}$, we expect spectral diffusion rates of $\SI{230}{ms}$ and  $\SI{91}{ms}$, respectively. For our experimentally determined $T_1$ time of $\SI{32.4}{s}$ we obtain $T_\mathrm{SD}=\SI{50}{ms}$. All of these estimates for $T_\mathrm{SD}$  exceed the observed $T_2$ times significantly and hence we expect that our $T_2$ measurements are not dominated by this mechanism. As the spin relaxation time represents an important parameter, we plan to investigate aspects of spin relaxation in these strongly coupled systems at a later stage in more detail using pulse sequences based on adiabatic pulses, optimal control pulses or a two-pulse saturation recovery, which have the potential to discern spectral diffusion from spin relaxation, excitation of a selected part of the spin ensemble and Purcell rates.

\section{Theoretical model}
In order to give a dynamical description of the echo trains we start from the inhomogeneous Tavis-Cummings Hamiltonian \cite{Tavis1968a}, 
\begin{eqnarray}
\mathcal{H}=&&\hbar\Delta_c\,a^\dag a +\frac{\hbar}{2}\sum_{j=1}^N\Delta_j \sigma_j^z
+\sum_{j=1}^N\hbar[g_j\sigma_j^-a^\dag+g_j^*\sigma_j^+a]\nonumber\\
&&+i\hbar[\eta(t) a^\dag-\eta^*(t) a],
\label{eq:Hamiltonian}
\end{eqnarray}
 where $\Delta_c\equiv \omega_c-\omega_p$ and $\Delta_j\equiv\omega_j-\omega_p$ are the detunings of the resonator frequency $\omega_c$ and of the individual spin frequencies $\omega_j$ from the frequency $\omega_p$ of the incoming driving pulse.
Here $a^\dag$ and $a$ are the creation and annihilation operators of the single resonator mode and $\sigma_j^z$, $\sigma_j^+$, and $\sigma_j^-$ are the Pauli operators corresponding to the individual spins. Without loss of generality we assume $\eta^*(t)=\eta(t)$ as well as $g^*_j=g_j$. The incoming driving pulse is characterized by the carrier frequency $\omega_p$ and the amplitude $\eta(t)$, which for simplicity is assumed to be of rectangular shape. Note that the Hamiltonian \eqref{eq:Hamiltonian} does not account for direct dipole-dipole interactions between the spins. Although dipole-dipole interactions do not seem to play a fundamental role in the formation of the echo pulses it would be interesting to investigate in future studies whether they have an impact on the shape of the echos.

A quantum master equation for the system's density matrix can be written as $d\rho/dt=-\frac{i}{\hbar}[\mathcal{H},\rho]+\mathcal{L}_D(\rho)$ \cite{CarmichaelQO2}, where $\mathcal{H}$ is the Hamiltonian \eqref{eq:Hamiltonian} and $\mathcal{L}_D(\rho)$ stands for the standard Lindblad superoperator
\begin{eqnarray}
\nonumber
&&\!\mathcal{L}_D(\rho)\!=\!\kappa\,(2a\rho a^\dagger-a^\dagger a\,\rho-\rho\, a^\dagger a)+\gamma_p\sum\limits_{j=1}^N(\sigma_j^z\rho\,\sigma_j^z-\rho\,)
\\[6pt]
&&+\gamma_h\sum\limits_{j=1}^N(2\sigma_j^-\rho\,\sigma_j^+-\sigma_j^+\sigma_j^-\rho-\rho\,\sigma_j^+\sigma_j^-).
\label{eq_Lindblad}
\end{eqnarray}
Here the first term describes the resonator losses with the decay rate $\kappa$ and the second and third term account for nonradiative and radiative dephasing of the individual spins characterized by the rates $\gamma_p$ and $\gamma_h$, respectively. Starting from the master equation given above, one can derive the equations of motion for the expectation value of any operator $O$ by $d\langle O\rangle/dt=\text{Tr}\{-\frac{i}{\hbar}[O,\mathcal{H}]\rho+O\mathcal{L}_D(\rho)\}$. In the limit of very large spin ensembles ($N\to\infty$), we can neglect correlations between the resonator field and individual spins ($\langle a^\dagger\sigma_j^-\rangle\approx\langle a^\dagger\rangle\langle \sigma_j^-\rangle$) \cite{Zens2019}. Thus, we obtain a closed set of first-order differential equations for the expectation values $\langle a\rangle$, $\langle \sigma_j^-\rangle$, and $\langle \sigma_j^z\rangle$, which is equivalent to the well-known Maxwell-Bloch equations: 
\begin{align} 
\label{eq:MaxwellBlocha}
&\frac{d}{dt}\langle a\rangle =-(\kappa+i\,\Delta_c)\langle a\rangle-i\,\sum_{j=1}^Ng_j\langle\sigma_j^-\rangle+\eta(t)\,, \\ 
&\frac{d}{dt}\langle \sigma_j^-\rangle =-(\gamma_\perp+i\,\Delta_j)\langle\sigma_j^-\rangle+i\,g_j\langle\sigma_j^z\rangle\langle a\rangle\,,\label{eq:MaxwellBlochb}\\[2mm]
&\frac{d}{dt}\langle\sigma_j^z\rangle =-\gamma_\parallel(\langle\sigma_j^z\rangle+1)
+2i\,g_j(\langle\sigma_j^-\rangle\langle a^\dag\rangle-\textit{c.c.})\,, 
\label{eq:MaxwellBlochc}
\end{align}  
with the resonator decay rate $\kappa$, the longitudinal spin relaxation rate $\gamma_\parallel=2\gamma_h=1/T_1$, and transverse spin relaxation rate $\gamma_\perp=\gamma_h+2\gamma_p=1/T_2$. 

As outlined in the main text, the spin ensemble is inhomogeneously broadened not only with regard to the individual spin frequencies $\omega_j$, but also through the coupling strengths $g_j$ due to the $B_1$ inhomogeneity. Since we are dealing with a sizable number of spins ($N\approx\num{1.06e10}$) inside the ensemble, the distributions of spin frequencies and couplings strengths are smooth functions around the mean values.
For simplicity, we assume in our calculations that all spins couple with the mean coupling strength $g_0=g_\mathrm{eff}/\sqrt{N}$ and we incorporate the inhomogeneaus broadening in a phenomenological Lorentzian spin spectral density 
\begin{equation}
\label{eq:Lorentzian}
\rho(\omega)=\frac{1}{\pi\gamma_s[1+(\frac{\omega-\omega_s}{\gamma_s})^2]}.
\end{equation} 
This frequency  distribution of spins is already sufficient to accurately describe the generation of multiple echo trains. Here $\gamma_s$ is the half width at half maximum and $\omega_s$ is the mean frequency of the spin distribution.

To solve \eqref{eq:MaxwellBlocha}-\eqref{eq:MaxwellBlochc} for the inhomogeneously broadened spin ensemble,  we discretize the phenomenological spin spectral density and divide the entire frequency range into $M=40001$ equidistant frequency clusters. Each cluster $k$ is then characterized by the mean coupling strength $g_0$, its detuning $\Delta_k=\omega_k-\omega_p$, and the number of spins inside this cluster. Eqs.~\ref{eq:MaxwellBlocha}-\ref{eq:MaxwellBlochc} can then be solved using a standard Runge-Kutta method.

Note that, along the lines of previous work \cite{Putz2017a,Krimer2019}, the distribution of coupling strengths $g_k$ can also be included in the phenomenological spin spectral density. Calculations using such a combined spin spectral density have also been carried out and showed qualitatively similar results. In order to obtain a full quantitative agreement between our theory and the experiment, however, the exact shape of the spectral spin and spatial coupling distribution has to be determined through extensive further theoretical and experimental work \cite{Sandner2012}. For reasons of clarity, we only present simulations in which the inhomogeneous broadening is included in the spin distribution alone, since these are already sufficient to describe the observed phenomenon of multiple echoes.

\section{A short review on multiple echo effects}
Multiple echo effects in nuclear and electron magnetic resonance (NMR, ESR) experiments have been observed in a number of experiments, although different underlying mechanism are presented. 
Multiple echo signatures were reported in NMR experiments of $^3$He, $^3$He/$^4$He mixtures as well as water\,\cite{Deville1979, Eska1981, Einzel1984, Bowtell1990}. In these experiments the occurrence of multiple echos is attributed to non-linear terms in the equation of motion governing the magnetization. In Fermi liquids, the non-linearity is introduced by the Leggett-Rice effect\,\cite{Einzel1984, Bedford1991a}. Neither effect plays a role in our experiments. Another source for non-linear terms in the Bloch equations is the dipolar demagnetizing field\,\cite{Bedford1991a}. The demagnetizing field is usually negligible NMR and ESR experiments, as it is suppressed by radiation damping\,\cite{Bowtell1996,Warren1995}. However, in the experiments presented in Ref.~\cite{Bowtell1996, Warren1995} a strong field gradient parallel to the static magnetic field was applied, which crucially alters the effect of the demagnetizing field on the dynamics\,\cite{Warren1995}. Another source of nonlinear spin dynamics is radiation damping\,\cite{Bloom1957, Augustine2002}. Radiation damping describes the effects of a backaction of the precessing spin magnetization on the RF coil or resonator, sharing some similarities with the strong coupling regime. Numerical simulations of the nonlinear Maxwell-Bloch equations indeed show the presence of multiple echos under certain conditions\,\cite{Vlassenbroek1995}.

The first occurrence of a multiple echo signal in ESR was reported by Gordon and Bowers\,\cite{Gordon1958}. Here, the authors conducted Hahn echo experiments of donors in silicon at a frequency of \SI{23}{GHz}. We are able to estimate the relevant coupling parameters from the information supplied in the text: Assuming a typical TE\textsubscript{102} cavity for operation at \SI{23}{GHz} and a sample volume of \SI{0.1}{cm^3}, we estimate a filling factor of $\approx\SI{6}{\%}$. We calculate the effective coupling rate using Eq.~\eqref{eq:geff} and a donor concentration of \SI{4e16}{cm^{-3}} and obtain $\geff\approx\SI{3.37}{MHz}$. With the spin relaxation rate $\gamma_s\approx\SI{560}{kHz}$ and the assumption of a moderate quality factor of $Q = 200$, we estimate a cooperativity of $C = 1.68$. Therefore, the occurrence of the second echo reported in Ref.~\cite{Gordon1958} can be in hindsight explained by our model.


\begin{thebibliography}{39}%
\makeatletter
\providecommand \@ifxundefined [1]{%
 \@ifx{#1\undefined}
}%
\providecommand \@ifnum [1]{%
 \ifnum #1\expandafter \@firstoftwo
 \else \expandafter \@secondoftwo
 \fi
}%
\providecommand \@ifx [1]{%
 \ifx #1\expandafter \@firstoftwo
 \else \expandafter \@secondoftwo
 \fi
}%
\providecommand \natexlab [1]{#1}%
\providecommand \enquote  [1]{``#1''}%
\providecommand \bibnamefont  [1]{#1}%
\providecommand \bibfnamefont [1]{#1}%
\providecommand \citenamefont [1]{#1}%
\providecommand \href@noop [0]{\@secondoftwo}%
\providecommand \href [0]{\begingroup \@sanitize@url \@href}%
\providecommand \@href[1]{\@@startlink{#1}\@@href}%
\providecommand \@@href[1]{\endgroup#1\@@endlink}%
\providecommand \@sanitize@url [0]{\catcode `\\12\catcode `\$12\catcode
  `\&12\catcode `\#12\catcode `\^12\catcode `\_12\catcode `\%12\relax}%
\providecommand \@@startlink[1]{}%
\providecommand \@@endlink[0]{}%
\providecommand \url  [0]{\begingroup\@sanitize@url \@url }%
\providecommand \@url [1]{\endgroup\@href {#1}{\urlprefix }}%
\providecommand \urlprefix  [0]{URL }%
\providecommand \Eprint [0]{\href }%
\providecommand \doibase [0]{http://dx.doi.org/}%
\providecommand \selectlanguage [0]{\@gobble}%
\providecommand \bibinfo  [0]{\@secondoftwo}%
\providecommand \bibfield  [0]{\@secondoftwo}%
\providecommand \translation [1]{[#1]}%
\providecommand \BibitemOpen [0]{}%
\providecommand \bibitemStop [0]{}%
\providecommand \bibitemNoStop [0]{.\EOS\space}%
\providecommand \EOS [0]{\spacefactor3000\relax}%
\providecommand \BibitemShut  [1]{\csname bibitem#1\endcsname}%
\let\auto@bib@innerbib\@empty
\bibitem [{\citenamefont {Prisner}\ \emph {et~al.}(2001)\citenamefont
  {Prisner}, \citenamefont {Rohrer},\ and\ \citenamefont
  {MacMillan}}]{Prisner2001}%
  \BibitemOpen
  \bibfield  {author} {\bibinfo {author} {\bibfnamefont {Thomas}\ \bibnamefont
  {Prisner}}, \bibinfo {author} {\bibfnamefont {Martin}\ \bibnamefont
  {Rohrer}}, \ and\ \bibinfo {author} {\bibfnamefont {Fraser}\ \bibnamefont
  {MacMillan}},\ }\bibfield  {title} {{\selectlanguage {english}\enquote
  {\bibinfo {title} {Pulsed {{EPR Spectroscopy}}: {{Biological
  Applications}}},}\ }}\href {\doibase 10.1146/annurev.physchem.52.1.279}
  {\bibfield  {journal} {\bibinfo  {journal} {Annual Review of Physical
  Chemistry}\ }\textbf {\bibinfo {volume} {52}},\ \bibinfo {pages} {279}
  (\bibinfo {year} {2001})}\BibitemShut {NoStop}%
\bibitem [{\citenamefont {Eaton}\ and\ \citenamefont
  {Eaton}(2015)}]{Eaton2015}%
  \BibitemOpen
  \bibfield  {author} {\bibinfo {author} {\bibfnamefont {Sandra~S.}\
  \bibnamefont {Eaton}}\ and\ \bibinfo {author} {\bibfnamefont {Gareth~R.}\
  \bibnamefont {Eaton}},\ }\bibfield  {title} {{\selectlanguage
  {english}\enquote {\bibinfo {title} {Multifrequency {{Pulsed EPR}} and the
  {{Characterization}} of {{Molecular Dynamics}}},}\ }}in\ \href {\doibase
  10.1016/bs.mie.2015.06.028} {{\selectlanguage {english}\emph {\bibinfo
  {booktitle} {Methods in {{Enzymology}}}}}},\ Vol.\ \bibinfo {volume} {563}\
  (\bibinfo  {publisher} {{Elsevier}},\ \bibinfo {year} {2015})\ p.~\bibinfo
  {pages} {37}\BibitemShut {NoStop}%
\bibitem [{\citenamefont {Baranov}\ \emph {et~al.}(2017)\citenamefont
  {Baranov}, \citenamefont {{von Bardeleben}}, \citenamefont {Jelezko},\ and\
  \citenamefont {Wrachtrup}}]{Baranov2017}%
  \BibitemOpen
  \bibfield  {author} {\bibinfo {author} {\bibfnamefont {Pavel~G.}\
  \bibnamefont {Baranov}}, \bibinfo {author} {\bibfnamefont {Hans~J\"urgen}\
  \bibnamefont {{von Bardeleben}}}, \bibinfo {author} {\bibfnamefont {Fedor}\
  \bibnamefont {Jelezko}}, \ and\ \bibinfo {author} {\bibfnamefont {J\"org}\
  \bibnamefont {Wrachtrup}},\ }\href {\doibase 10.1007/978-3-7091-1157-4}
  {{\bibinfo {title} {Magnetic {{Resonance}} of {{Semiconductors}} and
  {{Their Nanostructures}}}}},\ \bibinfo {series} {Springer {{Series}} in
  {{Materials Science}}}, Vol.\ \bibinfo {volume} {253}\ (\bibinfo  {publisher}
  {{Springer Vienna}},\ \bibinfo {address} {Vienna},\ \bibinfo {year}
  {2017})\BibitemShut {NoStop}%
\bibitem [{\citenamefont {Schirhagl}\ \emph {et~al.}(2014)\citenamefont
  {Schirhagl}, \citenamefont {Chang}, \citenamefont {Loretz},\ and\
  \citenamefont {Degen}}]{Schirhagl2014}%
  \BibitemOpen
  \bibfield  {author} {\bibinfo {author} {\bibfnamefont {Romana}\ \bibnamefont
  {Schirhagl}}, \bibinfo {author} {\bibfnamefont {Kevin}\ \bibnamefont
  {Chang}}, \bibinfo {author} {\bibfnamefont {Michael}\ \bibnamefont {Loretz}},
  \ and\ \bibinfo {author} {\bibfnamefont {Christian~L.}\ \bibnamefont
  {Degen}},\ }\bibfield  {title} {{\selectlanguage {english}\enquote {\bibinfo
  {title} {Nitrogen-{{Vacancy Centers}} in {{Diamond}}: {{Nanoscale Sensors}}
  for {{Physics}} and {{Biology}}},}\ }}\href {\doibase
  10.1146/annurev-physchem-040513-103659} {\bibfield  {journal} {\bibinfo
  {journal} {Annual Review of Physical Chemistry}\ }\textbf {\bibinfo {volume}
  {65}},\ \bibinfo {pages} {83} (\bibinfo {year} {2014})}\BibitemShut {NoStop}%
\bibitem [{\citenamefont {Devoret}\ and\ \citenamefont
  {Schoelkopf}(2013)}]{Devoret2013}%
  \BibitemOpen
  \bibfield  {author} {\bibinfo {author} {\bibfnamefont {M.~H.}\ \bibnamefont
  {Devoret}}\ and\ \bibinfo {author} {\bibfnamefont {R.~J.}\ \bibnamefont
  {Schoelkopf}},\ }\bibfield  {title} {{\selectlanguage {english}\enquote
  {\bibinfo {title} {Superconducting {{Circuits}} for {{Quantum Information}}:
  {{An Outlook}}},}\ }}\href {\doibase 10.1126/science.1231930} {\bibfield
  {journal} {\bibinfo  {journal} {Science}\ }\textbf {\bibinfo {volume}
  {339}},\ \bibinfo {pages} {1169} (\bibinfo {year} {2013})}\BibitemShut
  {NoStop}%
\bibitem [{\citenamefont {Zwanenburg}\ \emph {et~al.}(2013)\citenamefont
  {Zwanenburg}, \citenamefont {Dzurak}, \citenamefont {Morello}, \citenamefont
  {Simmons}, \citenamefont {Hollenberg}, \citenamefont {Klimeck}, \citenamefont
  {Rogge}, \citenamefont {Coppersmith},\ and\ \citenamefont
  {Eriksson}}]{Zwanenburg2013}%
  \BibitemOpen
  \bibfield  {author} {\bibinfo {author} {\bibfnamefont {Floris~A.}\
  \bibnamefont {Zwanenburg}}, \bibinfo {author} {\bibfnamefont {Andrew~S.}\
  \bibnamefont {Dzurak}}, \bibinfo {author} {\bibfnamefont {Andrea}\
  \bibnamefont {Morello}}, \bibinfo {author} {\bibfnamefont {Michelle~Y.}\
  \bibnamefont {Simmons}}, \bibinfo {author} {\bibfnamefont {Lloyd C.~L.}\
  \bibnamefont {Hollenberg}}, \bibinfo {author} {\bibfnamefont {Gerhard}\
  \bibnamefont {Klimeck}}, \bibinfo {author} {\bibfnamefont {Sven}\
  \bibnamefont {Rogge}}, \bibinfo {author} {\bibfnamefont {Susan~N.}\
  \bibnamefont {Coppersmith}}, \ and\ \bibinfo {author} {\bibfnamefont
  {Mark~A.}\ \bibnamefont {Eriksson}},\ }\bibfield  {title} {{\selectlanguage
  {english}\enquote {\bibinfo {title} {Silicon quantum electronics},}\ }}\href
  {\doibase 10.1103/RevModPhys.85.961} {\bibfield  {journal} {\bibinfo
  {journal} {Reviews of Modern Physics}\ }\textbf {\bibinfo {volume} {85}},\
  \bibinfo {pages} {961} (\bibinfo {year} {2013})}\BibitemShut {NoStop}%
\bibitem [{\citenamefont {Schweiger}\ and\ \citenamefont
  {Jeschke}(2001)}]{Schweiger2001}%
  \BibitemOpen
  \bibfield  {author} {\bibinfo {author} {\bibfnamefont {Arthur}\ \bibnamefont
  {Schweiger}}\ and\ \bibinfo {author} {\bibfnamefont {Gunnar}\ \bibnamefont
  {Jeschke}},\ }\href@noop {} {{\bibinfo {title} {Principles of Pulse
  Electron Paramagnetic Resonance}}}\ (\bibinfo  {publisher} {{Oxford
  University Press}},\ \bibinfo {address} {Oxford, UK ; New York},\ \bibinfo
  {year} {2001})\BibitemShut {NoStop}%
\bibitem [{\citenamefont {Hahn}(1950)}]{Hahn1950}%
  \BibitemOpen
  \bibfield  {author} {\bibinfo {author} {\bibfnamefont {E.~L.}\ \bibnamefont
  {Hahn}},\ }\bibfield  {title} {{\selectlanguage {english}\enquote {\bibinfo
  {title} {Spin {{Echoes}}},}\ }}\href {\doibase 10.1103/PhysRev.80.580}
  {\bibfield  {journal} {\bibinfo  {journal} {Physical Review}\ }\textbf
  {\bibinfo {volume} {80}},\ \bibinfo {pages} {580} (\bibinfo {year}
  {1950})}\BibitemShut {NoStop}%
\bibitem [{\citenamefont {Kubo}\ \emph {et~al.}(2010)\citenamefont {Kubo},
  \citenamefont {Ong}, \citenamefont {Bertet}, \citenamefont {Vion},
  \citenamefont {Jacques}, \citenamefont {Zheng}, \citenamefont {Dr\'eau},
  \citenamefont {Roch}, \citenamefont {Auffeves}, \citenamefont {Jelezko},
  \citenamefont {Wrachtrup}, \citenamefont {Barthe}, \citenamefont {Bergonzo},\
  and\ \citenamefont {Esteve}}]{Kubo2010}%
  \BibitemOpen
  \bibfield  {author} {\bibinfo {author} {\bibfnamefont {Y.}~\bibnamefont
  {Kubo}}, \bibinfo {author} {\bibfnamefont {F.~R.}\ \bibnamefont {Ong}},
  \bibinfo {author} {\bibfnamefont {P.}~\bibnamefont {Bertet}}, \bibinfo
  {author} {\bibfnamefont {D.}~\bibnamefont {Vion}}, \bibinfo {author}
  {\bibfnamefont {V.}~\bibnamefont {Jacques}}, \bibinfo {author} {\bibfnamefont
  {D.}~\bibnamefont {Zheng}}, \bibinfo {author} {\bibfnamefont
  {A.}~\bibnamefont {Dr\'eau}}, \bibinfo {author} {\bibfnamefont {J.-F.}\
  \bibnamefont {Roch}}, \bibinfo {author} {\bibfnamefont {A.}~\bibnamefont
  {Auffeves}}, \bibinfo {author} {\bibfnamefont {F.}~\bibnamefont {Jelezko}},
  \bibinfo {author} {\bibfnamefont {J.}~\bibnamefont {Wrachtrup}}, \bibinfo
  {author} {\bibfnamefont {M.~F.}\ \bibnamefont {Barthe}}, \bibinfo {author}
  {\bibfnamefont {P.}~\bibnamefont {Bergonzo}}, \ and\ \bibinfo {author}
  {\bibfnamefont {D.}~\bibnamefont {Esteve}},\ }\bibfield  {title}
  {{\selectlanguage {english}\enquote {\bibinfo {title} {Strong {{Coupling}} of
  a {{Spin Ensemble}} to a {{Superconducting Resonator}}},}\ }}\href {\doibase
  10.1103/PhysRevLett.105.140502} {\bibfield  {journal} {\bibinfo  {journal}
  {Physical Review Letters}\ }\textbf {\bibinfo {volume} {105}},\ \bibinfo
  {pages} {140502} (\bibinfo {year} {2010})}\BibitemShut {NoStop}%
\bibitem [{\citenamefont {Schuster}\ \emph {et~al.}(2010)\citenamefont
  {Schuster}, \citenamefont {Sears}, \citenamefont {Ginossar}, \citenamefont
  {DiCarlo}, \citenamefont {Frunzio}, \citenamefont {Morton}, \citenamefont
  {Wu}, \citenamefont {Briggs}, \citenamefont {Buckley}, \citenamefont
  {Awschalom},\ and\ \citenamefont {Schoelkopf}}]{Schuster2010}%
  \BibitemOpen
  \bibfield  {author} {\bibinfo {author} {\bibfnamefont {D.~I.}\ \bibnamefont
  {Schuster}}, \bibinfo {author} {\bibfnamefont {A.~P.}\ \bibnamefont {Sears}},
  \bibinfo {author} {\bibfnamefont {E.}~\bibnamefont {Ginossar}}, \bibinfo
  {author} {\bibfnamefont {L.}~\bibnamefont {DiCarlo}}, \bibinfo {author}
  {\bibfnamefont {L.}~\bibnamefont {Frunzio}}, \bibinfo {author} {\bibfnamefont
  {J.~J.~L.}\ \bibnamefont {Morton}}, \bibinfo {author} {\bibfnamefont
  {H.}~\bibnamefont {Wu}}, \bibinfo {author} {\bibfnamefont {G.~A.~D.}\
  \bibnamefont {Briggs}}, \bibinfo {author} {\bibfnamefont {B.~B.}\
  \bibnamefont {Buckley}}, \bibinfo {author} {\bibfnamefont {D.~D.}\
  \bibnamefont {Awschalom}}, \ and\ \bibinfo {author} {\bibfnamefont {R.~J.}\
  \bibnamefont {Schoelkopf}},\ }\bibfield  {title} {{\selectlanguage
  {english}\enquote {\bibinfo {title} {High-{{Cooperativity Coupling}} of
  {{Electron}}-{{Spin Ensembles}} to {{Superconducting Cavities}}},}\ }}\href
  {\doibase 10.1103/PhysRevLett.105.140501} {\bibfield  {journal} {\bibinfo
  {journal} {Physical Review Letters}\ }\textbf {\bibinfo {volume} {105}},\
  \bibinfo {pages} {140501} (\bibinfo {year} {2010})}\BibitemShut {NoStop}%
\bibitem [{\citenamefont {Ams\"uss}\ \emph {et~al.}(2011)\citenamefont
  {Ams\"uss}, \citenamefont {Koller}, \citenamefont {N\"obauer}, \citenamefont
  {Putz}, \citenamefont {Rotter}, \citenamefont {Sandner}, \citenamefont
  {Schneider}, \citenamefont {Schramb\"ock}, \citenamefont {Steinhauser},
  \citenamefont {Ritsch}, \citenamefont {Schmiedmayer},\ and\ \citenamefont
  {Majer}}]{Amsuss2011}%
  \BibitemOpen
  \bibfield  {author} {\bibinfo {author} {\bibfnamefont {R.}~\bibnamefont
  {Ams\"uss}}, \bibinfo {author} {\bibfnamefont {Ch.}\ \bibnamefont {Koller}},
  \bibinfo {author} {\bibfnamefont {T.}~\bibnamefont {N\"obauer}}, \bibinfo
  {author} {\bibfnamefont {S.}~\bibnamefont {Putz}}, \bibinfo {author}
  {\bibfnamefont {S.}~\bibnamefont {Rotter}}, \bibinfo {author} {\bibfnamefont
  {K.}~\bibnamefont {Sandner}}, \bibinfo {author} {\bibfnamefont
  {S.}~\bibnamefont {Schneider}}, \bibinfo {author} {\bibfnamefont
  {M.}~\bibnamefont {Schramb\"ock}}, \bibinfo {author} {\bibfnamefont
  {G.}~\bibnamefont {Steinhauser}}, \bibinfo {author} {\bibfnamefont
  {H.}~\bibnamefont {Ritsch}}, \bibinfo {author} {\bibfnamefont
  {J.}~\bibnamefont {Schmiedmayer}}, \ and\ \bibinfo {author} {\bibfnamefont
  {J.}~\bibnamefont {Majer}},\ }\bibfield  {title} {{\selectlanguage
  {english}\enquote {\bibinfo {title} {Cavity {{QED}} with {{Magnetically
  Coupled Collective Spin States}}},}\ }}\href {\doibase
  10.1103/PhysRevLett.107.060502} {\bibfield  {journal} {\bibinfo  {journal}
  {Physical Review Letters}\ }\textbf {\bibinfo {volume} {107}},\ \bibinfo
  {pages} {060502} (\bibinfo {year} {2011})}\BibitemShut {NoStop}%
\bibitem [{\citenamefont {Probst}\ \emph {et~al.}(2013)\citenamefont {Probst},
  \citenamefont {Rotzinger}, \citenamefont {W\"unsch}, \citenamefont {Jung},
  \citenamefont {Jerger}, \citenamefont {Siegel}, \citenamefont {Ustinov},\
  and\ \citenamefont {Bushev}}]{Probst2013}%
  \BibitemOpen
  \bibfield  {author} {\bibinfo {author} {\bibfnamefont {S.}~\bibnamefont
  {Probst}}, \bibinfo {author} {\bibfnamefont {H.}~\bibnamefont {Rotzinger}},
  \bibinfo {author} {\bibfnamefont {S.}~\bibnamefont {W\"unsch}}, \bibinfo
  {author} {\bibfnamefont {P.}~\bibnamefont {Jung}}, \bibinfo {author}
  {\bibfnamefont {M.}~\bibnamefont {Jerger}}, \bibinfo {author} {\bibfnamefont
  {M.}~\bibnamefont {Siegel}}, \bibinfo {author} {\bibfnamefont {A.~V.}\
  \bibnamefont {Ustinov}}, \ and\ \bibinfo {author} {\bibfnamefont {P.~A.}\
  \bibnamefont {Bushev}},\ }\bibfield  {title} {{\selectlanguage
  {english}\enquote {\bibinfo {title} {Anisotropic {{Rare}}-{{Earth Spin
  Ensemble Strongly Coupled}} to a {{Superconducting Resonator}}},}\ }}\href
  {\doibase 10.1103/PhysRevLett.110.157001} {\bibfield  {journal} {\bibinfo
  {journal} {Physical Review Letters}\ }\textbf {\bibinfo {volume} {110}},\
  \bibinfo {pages} {157001} (\bibinfo {year} {2013})}\BibitemShut {NoStop}%
\bibitem [{\citenamefont {Putz}\ \emph {et~al.}(2014)\citenamefont {Putz},
  \citenamefont {Krimer}, \citenamefont {Ams\"uss}, \citenamefont {Valookaran},
  \citenamefont {N\"obauer}, \citenamefont {Schmiedmayer}, \citenamefont
  {Rotter},\ and\ \citenamefont {Majer}}]{Putz2014}%
  \BibitemOpen
  \bibfield  {author} {\bibinfo {author} {\bibfnamefont {S.}~\bibnamefont
  {Putz}}, \bibinfo {author} {\bibfnamefont {D.~O.}\ \bibnamefont {Krimer}},
  \bibinfo {author} {\bibfnamefont {R.}~\bibnamefont {Ams\"uss}}, \bibinfo
  {author} {\bibfnamefont {A.}~\bibnamefont {Valookaran}}, \bibinfo {author}
  {\bibfnamefont {T.}~\bibnamefont {N\"obauer}}, \bibinfo {author}
  {\bibfnamefont {J.}~\bibnamefont {Schmiedmayer}}, \bibinfo {author}
  {\bibfnamefont {S.}~\bibnamefont {Rotter}}, \ and\ \bibinfo {author}
  {\bibfnamefont {J.}~\bibnamefont {Majer}},\ }\bibfield  {title} {\enquote
  {\bibinfo {title} {Protecting a spin ensemble against decoherence in the
  strong-coupling regime of cavity {{QED}}},}\ }\href {\doibase
  10.1038/nphys3050} {\bibfield  {journal} {\bibinfo  {journal} {Nature
  Physics}\ }\textbf {\bibinfo {volume} {10}},\ \bibinfo {pages} {720}
  (\bibinfo {year} {2014})}\BibitemShut {NoStop}%
\bibitem [{\citenamefont {Zollitsch}\ \emph {et~al.}(2015)\citenamefont
  {Zollitsch}, \citenamefont {Mueller}, \citenamefont {Franke}, \citenamefont
  {Goennenwein}, \citenamefont {Brandt}, \citenamefont {Gross},\ and\
  \citenamefont {Huebl}}]{Zollitsch2015}%
  \BibitemOpen
  \bibfield  {author} {\bibinfo {author} {\bibfnamefont {Christoph~W.}\
  \bibnamefont {Zollitsch}}, \bibinfo {author} {\bibfnamefont {Kai}\
  \bibnamefont {Mueller}}, \bibinfo {author} {\bibfnamefont {David~P.}\
  \bibnamefont {Franke}}, \bibinfo {author} {\bibfnamefont {Sebastian T.~B.}\
  \bibnamefont {Goennenwein}}, \bibinfo {author} {\bibfnamefont {Martin~S.}\
  \bibnamefont {Brandt}}, \bibinfo {author} {\bibfnamefont {Rudolf}\
  \bibnamefont {Gross}}, \ and\ \bibinfo {author} {\bibfnamefont {Hans}\
  \bibnamefont {Huebl}},\ }\bibfield  {title} {{\selectlanguage
  {english}\enquote {\bibinfo {title} {High cooperativity coupling between a
  phosphorus donor spin ensemble and a superconducting microwave resonator},}\
  }}\href {\doibase 10.1063/1.4932658} {\bibfield  {journal} {\bibinfo
  {journal} {Applied Physics Letters}\ }\textbf {\bibinfo {volume} {107}},\
  \bibinfo {pages} {142105} (\bibinfo {year} {2015})}\BibitemShut {NoStop}%
\bibitem [{\citenamefont {Morton}\ \emph {et~al.}(2008)\citenamefont {Morton},
  \citenamefont {Tyryshkin}, \citenamefont {Brown}, \citenamefont {Shankar},
  \citenamefont {Lovett}, \citenamefont {Ardavan}, \citenamefont {Schenkel},
  \citenamefont {Haller}, \citenamefont {Ager},\ and\ \citenamefont
  {Lyon}}]{Morton2008}%
  \BibitemOpen
  \bibfield  {author} {\bibinfo {author} {\bibfnamefont {John J.~L.}\
  \bibnamefont {Morton}}, \bibinfo {author} {\bibfnamefont {Alexei~M.}\
  \bibnamefont {Tyryshkin}}, \bibinfo {author} {\bibfnamefont {Richard~M.}\
  \bibnamefont {Brown}}, \bibinfo {author} {\bibfnamefont {Shyam}\ \bibnamefont
  {Shankar}}, \bibinfo {author} {\bibfnamefont {Brendon~W.}\ \bibnamefont
  {Lovett}}, \bibinfo {author} {\bibfnamefont {Arzhang}\ \bibnamefont
  {Ardavan}}, \bibinfo {author} {\bibfnamefont {Thomas}\ \bibnamefont
  {Schenkel}}, \bibinfo {author} {\bibfnamefont {Eugene~E.}\ \bibnamefont
  {Haller}}, \bibinfo {author} {\bibfnamefont {Joel~W.}\ \bibnamefont {Ager}},
  \ and\ \bibinfo {author} {\bibfnamefont {S.~A.}\ \bibnamefont {Lyon}},\
  }\bibfield  {title} {{\selectlanguage {english}\enquote {\bibinfo {title}
  {Solid-state quantum memory using the $^{31}${{P}} nuclear spin},}\ }}\href
  {\doibase 10.1038/nature07295} {\bibfield  {journal} {\bibinfo  {journal}
  {Nature}\ }\textbf {\bibinfo {volume} {455}},\ \bibinfo {pages} {1085}
  (\bibinfo {year} {2008})}\BibitemShut {NoStop}%
\bibitem [{\citenamefont {Bushev}\ \emph {et~al.}(2011)\citenamefont {Bushev},
  \citenamefont {Feofanov}, \citenamefont {Rotzinger}, \citenamefont
  {Protopopov}, \citenamefont {Cole}, \citenamefont {Wilson}, \citenamefont
  {Fischer}, \citenamefont {Lukashenko},\ and\ \citenamefont
  {Ustinov}}]{Bushev2011}%
  \BibitemOpen
  \bibfield  {author} {\bibinfo {author} {\bibfnamefont {P.}~\bibnamefont
  {Bushev}}, \bibinfo {author} {\bibfnamefont {A.~K.}\ \bibnamefont
  {Feofanov}}, \bibinfo {author} {\bibfnamefont {H.}~\bibnamefont {Rotzinger}},
  \bibinfo {author} {\bibfnamefont {I.}~\bibnamefont {Protopopov}}, \bibinfo
  {author} {\bibfnamefont {J.~H.}\ \bibnamefont {Cole}}, \bibinfo {author}
  {\bibfnamefont {C.~M.}\ \bibnamefont {Wilson}}, \bibinfo {author}
  {\bibfnamefont {G.}~\bibnamefont {Fischer}}, \bibinfo {author} {\bibfnamefont
  {A.}~\bibnamefont {Lukashenko}}, \ and\ \bibinfo {author} {\bibfnamefont
  {A.~V.}\ \bibnamefont {Ustinov}},\ }\bibfield  {title} {{\selectlanguage
  {english}\enquote {\bibinfo {title} {Ultralow-power spectroscopy of a
  rare-earth spin ensemble using a superconducting resonator},}\ }}\href
  {\doibase 10.1103/PhysRevB.84.060501} {\bibfield  {journal} {\bibinfo
  {journal} {Physical Review B}\ }\textbf {\bibinfo {volume} {84}},\ \bibinfo
  {pages} {060501(R)} (\bibinfo {year} {2011})}\BibitemShut {NoStop}%
\bibitem [{\citenamefont {Grezes}\ \emph {et~al.}(2016)\citenamefont {Grezes},
  \citenamefont {Kubo}, \citenamefont {Julsgaard}, \citenamefont {Umeda},
  \citenamefont {Isoya}, \citenamefont {Sumiya}, \citenamefont {Abe},
  \citenamefont {Onoda}, \citenamefont {Ohshima}, \citenamefont {Nakamura},
  \citenamefont {Diniz}, \citenamefont {Auffeves}, \citenamefont {Jacques},
  \citenamefont {Roch}, \citenamefont {Vion}, \citenamefont {Esteve},
  \citenamefont {Moelmer},\ and\ \citenamefont {Bertet}}]{Grezes2016}%
  \BibitemOpen
  \bibfield  {author} {\bibinfo {author} {\bibfnamefont {C\'ecile}\
  \bibnamefont {Grezes}}, \bibinfo {author} {\bibfnamefont {Yuimaru}\
  \bibnamefont {Kubo}}, \bibinfo {author} {\bibfnamefont {Brian}\ \bibnamefont
  {Julsgaard}}, \bibinfo {author} {\bibfnamefont {Takahide}\ \bibnamefont
  {Umeda}}, \bibinfo {author} {\bibfnamefont {Junichi}\ \bibnamefont {Isoya}},
  \bibinfo {author} {\bibfnamefont {Hitoshi}\ \bibnamefont {Sumiya}}, \bibinfo
  {author} {\bibfnamefont {Hiroshi}\ \bibnamefont {Abe}}, \bibinfo {author}
  {\bibfnamefont {Shinobu}\ \bibnamefont {Onoda}}, \bibinfo {author}
  {\bibfnamefont {Takeshi}\ \bibnamefont {Ohshima}}, \bibinfo {author}
  {\bibfnamefont {Kazuo}\ \bibnamefont {Nakamura}}, \bibinfo {author}
  {\bibfnamefont {Igor}\ \bibnamefont {Diniz}}, \bibinfo {author}
  {\bibfnamefont {Alexia}\ \bibnamefont {Auffeves}}, \bibinfo {author}
  {\bibfnamefont {Vincent}\ \bibnamefont {Jacques}}, \bibinfo {author}
  {\bibfnamefont {Jean-Fran{\c c}ois}\ \bibnamefont {Roch}}, \bibinfo {author}
  {\bibfnamefont {Denis}\ \bibnamefont {Vion}}, \bibinfo {author}
  {\bibfnamefont {Daniel}\ \bibnamefont {Esteve}}, \bibinfo {author}
  {\bibfnamefont {Klaus}\ \bibnamefont {Moelmer}}, \ and\ \bibinfo {author}
  {\bibfnamefont {Patrice}\ \bibnamefont {Bertet}},\ }\bibfield  {title}
  {{\selectlanguage {english}\enquote {\bibinfo {title} {Towards a
  spin-ensemble quantum memory for superconducting qubits},}\ }}\href {\doibase
  10.1016/j.crhy.2016.07.006} {\bibfield  {journal} {\bibinfo  {journal}
  {Comptes Rendus Physique}\ }\textbf {\bibinfo {volume} {17}},\ \bibinfo
  {pages} {693} (\bibinfo {year} {2016})}\BibitemShut {NoStop}%
\bibitem [{\citenamefont {Bienfait}\ \emph {et~al.}(2016)\citenamefont
  {Bienfait}, \citenamefont {Pla}, \citenamefont {Kubo}, \citenamefont {Stern},
  \citenamefont {Zhou}, \citenamefont {Lo}, \citenamefont {Weis}, \citenamefont
  {Schenkel}, \citenamefont {Thewalt}, \citenamefont {Vion}, \citenamefont
  {Esteve}, \citenamefont {Julsgaard}, \citenamefont {M\o{}lmer}, \citenamefont
  {Morton},\ and\ \citenamefont {Bertet}}]{Bienfait2016}%
  \BibitemOpen
  \bibfield  {author} {\bibinfo {author} {\bibfnamefont {A.}~\bibnamefont
  {Bienfait}}, \bibinfo {author} {\bibfnamefont {J.~J.}\ \bibnamefont {Pla}},
  \bibinfo {author} {\bibfnamefont {Y.}~\bibnamefont {Kubo}}, \bibinfo {author}
  {\bibfnamefont {M.}~\bibnamefont {Stern}}, \bibinfo {author} {\bibfnamefont
  {X.}~\bibnamefont {Zhou}}, \bibinfo {author} {\bibfnamefont {C.~C.}\
  \bibnamefont {Lo}}, \bibinfo {author} {\bibfnamefont {C.~D.}\ \bibnamefont
  {Weis}}, \bibinfo {author} {\bibfnamefont {T.}~\bibnamefont {Schenkel}},
  \bibinfo {author} {\bibfnamefont {M.~L.~W.}\ \bibnamefont {Thewalt}},
  \bibinfo {author} {\bibfnamefont {D.}~\bibnamefont {Vion}}, \bibinfo {author}
  {\bibfnamefont {D.}~\bibnamefont {Esteve}}, \bibinfo {author} {\bibfnamefont
  {B.}~\bibnamefont {Julsgaard}}, \bibinfo {author} {\bibfnamefont
  {K.}~\bibnamefont {M\o{}lmer}}, \bibinfo {author} {\bibfnamefont {J.~J.~L.}\
  \bibnamefont {Morton}}, \ and\ \bibinfo {author} {\bibfnamefont
  {P.}~\bibnamefont {Bertet}},\ }\bibfield  {title} {{\selectlanguage
  {english}\enquote {\bibinfo {title} {Reaching the quantum limit of
  sensitivity in electron spin resonance},}\ }}\href {\doibase
  10.1038/nnano.2015.282} {\bibfield  {journal} {\bibinfo  {journal} {Nature
  Nanotechnology}\ }\textbf {\bibinfo {volume} {11}},\ \bibinfo {pages} {253}
  (\bibinfo {year} {2016})}\BibitemShut {NoStop}%
\bibitem [{\citenamefont {Eichler}\ \emph {et~al.}(2017)\citenamefont
  {Eichler}, \citenamefont {Sigillito}, \citenamefont {Lyon},\ and\
  \citenamefont {Petta}}]{Eichler2017}%
  \BibitemOpen
  \bibfield  {author} {\bibinfo {author} {\bibfnamefont {C.}~\bibnamefont
  {Eichler}}, \bibinfo {author} {\bibfnamefont {A.~J.}\ \bibnamefont
  {Sigillito}}, \bibinfo {author} {\bibfnamefont {S.~A.}\ \bibnamefont {Lyon}},
  \ and\ \bibinfo {author} {\bibfnamefont {J.~R.}\ \bibnamefont {Petta}},\
  }\bibfield  {title} {{\selectlanguage {english}\enquote {\bibinfo {title}
  {Electron {{Spin Resonance}} at the {{Level}} of $10^4$ {{Spins Using Low
  Impedance Superconducting Resonators}}},}\ }}\href {\doibase
  10.1103/PhysRevLett.118.037701} {\bibfield  {journal} {\bibinfo  {journal}
  {Physical Review Letters}\ }\textbf {\bibinfo {volume} {118}},\ \bibinfo
  {pages} {037701} (\bibinfo {year} {2017})}\BibitemShut {NoStop}%
\bibitem [{\citenamefont {Levitt}(2008)}]{Levitt2008}%
  \BibitemOpen
  \bibfield  {author} {\bibinfo {author} {\bibfnamefont {Malcolm~H.}\
  \bibnamefont {Levitt}},\ }\href@noop {} { {\bibinfo {title} {Spin
  Dynamics: Basics of Nuclear Magnetic Resonance}}},\ \bibinfo {edition} {2nd}\
  ed.\ (\bibinfo  {publisher} {{John Wiley \& Sons}},\ \bibinfo {address}
  {Chichester, England ; Hoboken, NJ},\ \bibinfo {year} {2008})\BibitemShut
  {NoStop}%
\bibitem [{\citenamefont {Rose}\ \emph {et~al.}(2017)\citenamefont {Rose},
  \citenamefont {Tyryshkin}, \citenamefont {Riemann}, \citenamefont
  {Abrosimov}, \citenamefont {Becker}, \citenamefont {Pohl}, \citenamefont
  {Thewalt}, \citenamefont {Itoh},\ and\ \citenamefont {Lyon}}]{Rose2017}%
  \BibitemOpen
  \bibfield  {author} {\bibinfo {author} {\bibfnamefont {B.~C.}\ \bibnamefont
  {Rose}}, \bibinfo {author} {\bibfnamefont {A.~M.}\ \bibnamefont {Tyryshkin}},
  \bibinfo {author} {\bibfnamefont {H.}~\bibnamefont {Riemann}}, \bibinfo
  {author} {\bibfnamefont {N.~V.}\ \bibnamefont {Abrosimov}}, \bibinfo {author}
  {\bibfnamefont {P.}~\bibnamefont {Becker}}, \bibinfo {author} {\bibfnamefont
  {H.-J.}\ \bibnamefont {Pohl}}, \bibinfo {author} {\bibfnamefont {M.~L.~W.}\
  \bibnamefont {Thewalt}}, \bibinfo {author} {\bibfnamefont {K.~M.}\
  \bibnamefont {Itoh}}, \ and\ \bibinfo {author} {\bibfnamefont {S.~A.}\
  \bibnamefont {Lyon}},\ }\bibfield  {title} {{\selectlanguage
  {english}\enquote {\bibinfo {title} {Coherent {{Rabi Dynamics}} of a
  {{Superradiant Spin Ensemble}} in a {{Microwave Cavity}}},}\ }}\href
  {\doibase 10.1103/PhysRevX.7.031002} {\bibfield  {journal} {\bibinfo
  {journal} {Physical Review X}\ }\textbf {\bibinfo {volume} {7}},\ \bibinfo
  {pages} {031002} (\bibinfo {year} {2017})}\BibitemShut {NoStop}%
\bibitem [{\citenamefont {Putz}\ \emph {et~al.}(2017)\citenamefont {Putz},
  \citenamefont {Angerer}, \citenamefont {Krimer}, \citenamefont {Glattauer},
  \citenamefont {Munro}, \citenamefont {Rotter}, \citenamefont {Schmiedmayer},\
  and\ \citenamefont {Majer}}]{Putz2017a}%
  \BibitemOpen
  \bibfield  {author} {\bibinfo {author} {\bibfnamefont {Stefan}\ \bibnamefont
  {Putz}}, \bibinfo {author} {\bibfnamefont {Andreas}\ \bibnamefont {Angerer}},
  \bibinfo {author} {\bibfnamefont {Dmitry~O.}\ \bibnamefont {Krimer}},
  \bibinfo {author} {\bibfnamefont {Ralph}\ \bibnamefont {Glattauer}}, \bibinfo
  {author} {\bibfnamefont {William~J.}\ \bibnamefont {Munro}}, \bibinfo
  {author} {\bibfnamefont {Stefan}\ \bibnamefont {Rotter}}, \bibinfo {author}
  {\bibfnamefont {J\"org}\ \bibnamefont {Schmiedmayer}}, \ and\ \bibinfo
  {author} {\bibfnamefont {Johannes}\ \bibnamefont {Majer}},\ }\bibfield
  {title} {{\selectlanguage {english}\enquote {\bibinfo {title} {Spectral hole
  burning and its application in microwave photonics},}\ }}\href {\doibase
  10.1038/nphoton.2016.225} {\bibfield  {journal} {\bibinfo  {journal} {Nature
  Photonics}\ }\textbf {\bibinfo {volume} {11}},\ \bibinfo {pages} {36}
  (\bibinfo {year} {2017})}\BibitemShut {NoStop}%
\bibitem [{\citenamefont {Angerer}\ \emph {et~al.}(2017)\citenamefont
  {Angerer}, \citenamefont {Putz}, \citenamefont {Krimer}, \citenamefont
  {Astner}, \citenamefont {Zens}, \citenamefont {Glattauer}, \citenamefont
  {Streltsov}, \citenamefont {Munro}, \citenamefont {Nemoto}, \citenamefont
  {Rotter}, \citenamefont {Schmiedmayer},\ and\ \citenamefont
  {Majer}}]{Angerer2017}%
  \BibitemOpen
  \bibfield  {author} {\bibinfo {author} {\bibfnamefont {Andreas}\ \bibnamefont
  {Angerer}}, \bibinfo {author} {\bibfnamefont {Stefan}\ \bibnamefont {Putz}},
  \bibinfo {author} {\bibfnamefont {Dmitry~O.}\ \bibnamefont {Krimer}},
  \bibinfo {author} {\bibfnamefont {Thomas}\ \bibnamefont {Astner}}, \bibinfo
  {author} {\bibfnamefont {Matthias}\ \bibnamefont {Zens}}, \bibinfo {author}
  {\bibfnamefont {Ralph}\ \bibnamefont {Glattauer}}, \bibinfo {author}
  {\bibfnamefont {Kirill}\ \bibnamefont {Streltsov}}, \bibinfo {author}
  {\bibfnamefont {William~J.}\ \bibnamefont {Munro}}, \bibinfo {author}
  {\bibfnamefont {Kae}\ \bibnamefont {Nemoto}}, \bibinfo {author}
  {\bibfnamefont {Stefan}\ \bibnamefont {Rotter}}, \bibinfo {author}
  {\bibfnamefont {J\"org}\ \bibnamefont {Schmiedmayer}}, \ and\ \bibinfo
  {author} {\bibfnamefont {Johannes}\ \bibnamefont {Majer}},\ }\bibfield
  {title} {{\selectlanguage {english}\enquote {\bibinfo {title} {Ultralong
  relaxation times in bistable hybrid quantum systems},}\ }}\href {\doibase
  10.1126/sciadv.1701626} {\bibfield  {journal} {\bibinfo  {journal} {Science
  Advances}\ }\textbf {\bibinfo {volume} {3}},\ \bibinfo {pages} {e1701626}
  (\bibinfo {year} {2017})}\BibitemShut {NoStop}%
\bibitem [{\citenamefont {Gordon}\ and\ \citenamefont
  {Bowers}(1958)}]{Gordon1958}%
  \BibitemOpen
  \bibfield  {author} {\bibinfo {author} {\bibfnamefont {J.~P.}\ \bibnamefont
  {Gordon}}\ and\ \bibinfo {author} {\bibfnamefont {K.~D.}\ \bibnamefont
  {Bowers}},\ }\bibfield  {title} {{\selectlanguage {english}\enquote {\bibinfo
  {title} {Microwave {{Spin Echoes}} from {{Donor Electrons}} in
  {{Silicon}}},}\ }}\href {\doibase 10.1103/PhysRevLett.1.368} {\bibfield
  {journal} {\bibinfo  {journal} {Physical Review Letters}\ }\textbf {\bibinfo
  {volume} {1}},\ \bibinfo {pages} {368--370} (\bibinfo {year}
  {1958})}\BibitemShut {NoStop}%
\bibitem [{Sup()}]{SupplementaryInformation}%
  \BibitemOpen
  \href@noop {} {}\bibinfo {note} {See Supplemental Material at
  \url{https://...} for details on the experimental setup, the theoretical
  description as well as additional measurements, including spin life time and
  spin coherence time measurements.}\BibitemShut {Stop}%
\bibitem [{\citenamefont {Probst}\ \emph {et~al.}(2015)\citenamefont {Probst},
  \citenamefont {Song}, \citenamefont {Bushev}, \citenamefont {Ustinov},\ and\
  \citenamefont {Weides}}]{Probst2015}%
  \BibitemOpen
  \bibfield  {author} {\bibinfo {author} {\bibfnamefont {S.}~\bibnamefont
  {Probst}}, \bibinfo {author} {\bibfnamefont {F.~B.}\ \bibnamefont {Song}},
  \bibinfo {author} {\bibfnamefont {P.~A.}\ \bibnamefont {Bushev}}, \bibinfo
  {author} {\bibfnamefont {A.~V.}\ \bibnamefont {Ustinov}}, \ and\ \bibinfo
  {author} {\bibfnamefont {M.}~\bibnamefont {Weides}},\ }\bibfield  {title}
  {{\selectlanguage {english}\enquote {\bibinfo {title} {Efficient and robust
  analysis of complex scattering data under noise in microwave resonators},}\
  }}\href {\doibase 10.1063/1.4907935} {\bibfield  {journal} {\bibinfo
  {journal} {Review of Scientific Instruments}\ }\textbf {\bibinfo {volume}
  {86}},\ \bibinfo {pages} {024706} (\bibinfo {year} {2015})}\BibitemShut
  {NoStop}%
\bibitem [{\citenamefont {Weichselbaumer}\ \emph {et~al.}(2019)\citenamefont
  {Weichselbaumer}, \citenamefont {Natzkin}, \citenamefont {Zollitsch},
  \citenamefont {Weiler}, \citenamefont {Gross},\ and\ \citenamefont
  {Huebl}}]{Weichselbaumer2019}%
  \BibitemOpen
  \bibfield  {author} {\bibinfo {author} {\bibfnamefont {Stefan}\ \bibnamefont
  {Weichselbaumer}}, \bibinfo {author} {\bibfnamefont {Petio}\ \bibnamefont
  {Natzkin}}, \bibinfo {author} {\bibfnamefont {Christoph~W.}\ \bibnamefont
  {Zollitsch}}, \bibinfo {author} {\bibfnamefont {Mathias}\ \bibnamefont
  {Weiler}}, \bibinfo {author} {\bibfnamefont {Rudolf}\ \bibnamefont {Gross}},
  \ and\ \bibinfo {author} {\bibfnamefont {Hans}\ \bibnamefont {Huebl}},\
  }\bibfield  {title} {{\selectlanguage {english}\enquote {\bibinfo {title}
  {Quantitative {{Modeling}} of {{Superconducting Planar Resonators}} for
  {{Electron Spin Resonance}}},}\ }}\href {\doibase
  10.1103/PhysRevApplied.12.024021} {\bibfield  {journal} {\bibinfo  {journal}
  {Physical Review Applied}\ }\textbf {\bibinfo {volume} {12}},\ \bibinfo
  {pages} {024021} (\bibinfo {year} {2019})}\BibitemShut {NoStop}%
\bibitem [{\citenamefont {Poindexter}\ \emph {et~al.}(1981)\citenamefont
  {Poindexter}, \citenamefont {Caplan}, \citenamefont {Deal},\ and\
  \citenamefont {Razouk}}]{Poindexter1981}%
  \BibitemOpen
  \bibfield  {author} {\bibinfo {author} {\bibfnamefont {Edward~H.}\
  \bibnamefont {Poindexter}}, \bibinfo {author} {\bibfnamefont {Philip~J.}\
  \bibnamefont {Caplan}}, \bibinfo {author} {\bibfnamefont {Bruce~E.}\
  \bibnamefont {Deal}}, \ and\ \bibinfo {author} {\bibfnamefont {Reda~R.}\
  \bibnamefont {Razouk}},\ }\bibfield  {title} {{\selectlanguage
  {english}\enquote {\bibinfo {title} {Interface states and electron spin
  resonance centers in thermally oxidized (111) and (100) silicon wafers},}\
  }}\href {\doibase 10.1063/1.328771} {\bibfield  {journal} {\bibinfo
  {journal} {Journal of Applied Physics}\ }\textbf {\bibinfo {volume} {52}},\
  \bibinfo {pages} {879} (\bibinfo {year} {1981})}\BibitemShut {NoStop}%
\bibitem [{\citenamefont {Stesmans}\ and\ \citenamefont
  {Afanas'ev}(1998)}]{Stesmans1998}%
  \BibitemOpen
  \bibfield  {author} {\bibinfo {author} {\bibfnamefont {A.}~\bibnamefont
  {Stesmans}}\ and\ \bibinfo {author} {\bibfnamefont {V.~V.}\ \bibnamefont
  {Afanas'ev}},\ }\bibfield  {title} {{\selectlanguage {english}\enquote
  {\bibinfo {title} {Electron spin resonance features of interface defects in
  thermal (100){{Si}}/{{SiO}}$_2$},}\ }}\href {\doibase 10.1063/1.367005}
  {\bibfield  {journal} {\bibinfo  {journal} {Journal of Applied Physics}\
  }\textbf {\bibinfo {volume} {83}},\ \bibinfo {pages} {2449} (\bibinfo {year}
  {1998})}\BibitemShut {NoStop}%
\bibitem [{\citenamefont {Feher}\ \emph {et~al.}(1955)\citenamefont {Feher},
  \citenamefont {Fletcher},\ and\ \citenamefont {Gere}}]{Feher1955}%
  \BibitemOpen
  \bibfield  {author} {\bibinfo {author} {\bibfnamefont {G.}~\bibnamefont
  {Feher}}, \bibinfo {author} {\bibfnamefont {R.~C.}\ \bibnamefont {Fletcher}},
  \ and\ \bibinfo {author} {\bibfnamefont {E.~A.}\ \bibnamefont {Gere}},\
  }\bibfield  {title} {{\selectlanguage {english}\enquote {\bibinfo {title}
  {Exchange {{Effects}} in {{Spin Resonance}} of {{Impurity Atoms}} in
  {{Silicon}}},}\ }}\href {\doibase 10.1103/PhysRev.100.1784.2} {\bibfield
  {journal} {\bibinfo  {journal} {Physical Review}\ }\textbf {\bibinfo {volume}
  {100}},\ \bibinfo {pages} {1784} (\bibinfo {year} {1955})}\BibitemShut
  {NoStop}%
\bibitem [{\citenamefont {J\'erome}\ and\ \citenamefont
  {Winter}(1964)}]{Jerome1964}%
  \BibitemOpen
  \bibfield  {author} {\bibinfo {author} {\bibfnamefont {D.}~\bibnamefont
  {J\'erome}}\ and\ \bibinfo {author} {\bibfnamefont {J.~M.}\ \bibnamefont
  {Winter}},\ }\bibfield  {title} {{\selectlanguage {english}\enquote {\bibinfo
  {title} {Electron {{Spin Resonance}} on {{Interacting Donors}} in
  {{Silicon}}},}\ }}\href {\doibase 10.1103/PhysRev.134.A1001} {\bibfield
  {journal} {\bibinfo  {journal} {Physical Review}\ }\textbf {\bibinfo {volume}
  {134}},\ \bibinfo {pages} {A1001} (\bibinfo {year} {1964})}\BibitemShut
  {NoStop}%
\bibitem [{\citenamefont {Morigaki}\ and\ \citenamefont
  {Maekawa}(1972)}]{Morigaki1972}%
  \BibitemOpen
  \bibfield  {author} {\bibinfo {author} {\bibfnamefont {Kazuo}\ \bibnamefont
  {Morigaki}}\ and\ \bibinfo {author} {\bibfnamefont {Shigeru}\ \bibnamefont
  {Maekawa}},\ }\bibfield  {title} {{\selectlanguage {english}\enquote
  {\bibinfo {title} {Electron {{Spin Resonance Studies}} of {{Interacting Donor
  Clusters}} in {{Phosphorus}}-{{Doped Silicon}}},}\ }}\href {\doibase
  10.1143/JPSJ.32.462} {\bibfield  {journal} {\bibinfo  {journal} {Journal of
  the Physical Society of Japan}\ }\textbf {\bibinfo {volume} {32}},\ \bibinfo
  {pages} {462} (\bibinfo {year} {1972})}\BibitemShut {NoStop}%
\bibitem [{\citenamefont {Shankar}\ \emph {et~al.}(2015)\citenamefont
  {Shankar}, \citenamefont {Tyryshkin},\ and\ \citenamefont
  {Lyon}}]{Shankar2015}%
  \BibitemOpen
  \bibfield  {author} {\bibinfo {author} {\bibfnamefont {S.}~\bibnamefont
  {Shankar}}, \bibinfo {author} {\bibfnamefont {A.~M.}\ \bibnamefont
  {Tyryshkin}}, \ and\ \bibinfo {author} {\bibfnamefont {S.~A.}\ \bibnamefont
  {Lyon}},\ }\bibfield  {title} {{\selectlanguage {english}\enquote {\bibinfo
  {title} {{{ESR}} measurements of phosphorus dimers in isotopically enriched
  $^{28}$si silicon},}\ }}\href {\doibase 10.1103/PhysRevB.91.245206}
  {\bibfield  {journal} {\bibinfo  {journal} {Physical Review B}\ }\textbf
  {\bibinfo {volume} {91}},\ \bibinfo {pages} {245206} (\bibinfo {year}
  {2015})}\BibitemShut {NoStop}%
\bibitem [{\citenamefont {Herskind}\ \emph {et~al.}(2009)\citenamefont
  {Herskind}, \citenamefont {Dantan}, \citenamefont {Marler}, \citenamefont
  {Albert},\ and\ \citenamefont {Drewsen}}]{Herskind2009}%
  \BibitemOpen
  \bibfield  {author} {\bibinfo {author} {\bibfnamefont {Peter~F.}\
  \bibnamefont {Herskind}}, \bibinfo {author} {\bibfnamefont {Aur\'elien}\
  \bibnamefont {Dantan}}, \bibinfo {author} {\bibfnamefont {Joan~P.}\
  \bibnamefont {Marler}}, \bibinfo {author} {\bibfnamefont {Magnus}\
  \bibnamefont {Albert}}, \ and\ \bibinfo {author} {\bibfnamefont {Michael}\
  \bibnamefont {Drewsen}},\ }\bibfield  {title} {\enquote {\bibinfo {title}
  {Realization of collective strong coupling with ion {{Coulomb}} crystals in
  an optical cavity},}\ }\href {\doibase 10.1038/nphys1302} {\bibfield
  {journal} {\bibinfo  {journal} {Nature Physics}\ }\textbf {\bibinfo {volume}
  {5}},\ \bibinfo {pages} {494} (\bibinfo {year} {2009})}\BibitemShut {NoStop}%
\bibitem [{\citenamefont {Huebl}\ \emph {et~al.}(2013)\citenamefont {Huebl},
  \citenamefont {Zollitsch}, \citenamefont {Lotze}, \citenamefont {Hocke},
  \citenamefont {Greifenstein}, \citenamefont {Marx}, \citenamefont {Gross},\
  and\ \citenamefont {Goennenwein}}]{Huebl2013}%
  \BibitemOpen
  \bibfield  {author} {\bibinfo {author} {\bibfnamefont {Hans}\ \bibnamefont
  {Huebl}}, \bibinfo {author} {\bibfnamefont {Christoph~W.}\ \bibnamefont
  {Zollitsch}}, \bibinfo {author} {\bibfnamefont {Johannes}\ \bibnamefont
  {Lotze}}, \bibinfo {author} {\bibfnamefont {Fredrik}\ \bibnamefont {Hocke}},
  \bibinfo {author} {\bibfnamefont {Moritz}\ \bibnamefont {Greifenstein}},
  \bibinfo {author} {\bibfnamefont {Achim}\ \bibnamefont {Marx}}, \bibinfo
  {author} {\bibfnamefont {Rudolf}\ \bibnamefont {Gross}}, \ and\ \bibinfo
  {author} {\bibfnamefont {Sebastian T.~B.}\ \bibnamefont {Goennenwein}},\
  }\bibfield  {title} {{\selectlanguage {english}\enquote {\bibinfo {title}
  {High {{Cooperativity}} in {{Coupled Microwave Resonator Ferrimagnetic
  Insulator Hybrids}}},}\ }}\href {\doibase 10.1103/PhysRevLett.111.127003}
  {\bibfield  {journal} {\bibinfo  {journal} {Physical Review Letters}\
  }\textbf {\bibinfo {volume} {111}},\ \bibinfo {pages} {127003} (\bibinfo
  {year} {2013})}\BibitemShut {NoStop}%
\bibitem [{Note1()}]{Note1}%
  \BibitemOpen
  \bibinfo {note} {Note that we use for the P$_2$ dimer measurements the same
  measurement protocol as for hyperfine transitions. Within the noise budget, a
  second echo should be detectable, if its echo amplitudes scale in the same
  manner as for the strong coupling case. However, we do not observe such a
  subsequent echo.}\BibitemShut {Stop}%
\bibitem [{\citenamefont {Zens}\ \emph {et~al.}(2019)\citenamefont {Zens},
  \citenamefont {Krimer},\ and\ \citenamefont {Rotter}}]{Zens2019}%
  \BibitemOpen
  \bibfield  {author} {\bibinfo {author} {\bibfnamefont {M.}~\bibnamefont
  {Zens}}, \bibinfo {author} {\bibfnamefont {D.O.}\ \bibnamefont {Krimer}}, \
  and\ \bibinfo {author} {\bibfnamefont {S.}~\bibnamefont {Rotter}},\
  }\bibfield  {title} {\enquote {\bibinfo {title} {{Critical phenomena and
  nonlinear dynamics in a spin ensemble strongly coupled to a cavity. II.
  Semiclassical-to-quantum boundary}},}\ }\href {\doibase
  10.1103/PhysRevA.100.013856} {\bibfield  {journal} {\bibinfo  {journal}
  {Physical Review A}\ }\textbf {\bibinfo {volume} {100}},\ \bibinfo {pages}
  {013856} (\bibinfo {year} {2019})}\BibitemShut {NoStop}%
\bibitem [{\citenamefont {Krimer}\ \emph {et~al.}(2019)\citenamefont {Krimer},
  \citenamefont {Zens},\ and\ \citenamefont {Rotter}}]{Krimer2019}%
  \BibitemOpen
  \bibfield  {author} {\bibinfo {author} {\bibfnamefont {Dmitry~O.}\
  \bibnamefont {Krimer}}, \bibinfo {author} {\bibfnamefont {Matthias}\
  \bibnamefont {Zens}}, \ and\ \bibinfo {author} {\bibfnamefont {Stefan}\
  \bibnamefont {Rotter}},\ }\bibfield  {title} {\enquote {\bibinfo {title}
  {{Critical phenomena and nonlinear dynamics in a spin ensemble strongly
  coupled to a cavity. I. Semiclassical approach}},}\ }\href {\doibase
  10.1103/PhysRevA.100.013855} {\bibfield  {journal} {\bibinfo  {journal}
  {Physical Review A}\ }\textbf {\bibinfo {volume} {100}},\ \bibinfo {pages}
  {013855} (\bibinfo {year} {2019})}\BibitemShut {NoStop}%
\bibitem [{\citenamefont {Debnath}\ \emph {et~al.}(2020)\citenamefont
  {Debnath}, \citenamefont {Dold}, \citenamefont {Morton},\ and\ \citenamefont
  {M{\o}lmer}}]{Debnath:2020}%
  \BibitemOpen
  \bibfield  {author} {\bibinfo {author} {\bibfnamefont {Kamanasish}\
  \bibnamefont {Debnath}}, \bibinfo {author} {\bibfnamefont {Gavin}\
  \bibnamefont {Dold}}, \bibinfo {author} {\bibfnamefont {John J~L}\
  \bibnamefont {Morton}}, \ and\ \bibinfo {author} {\bibfnamefont {Klaus}\
  \bibnamefont {M{\o}lmer}},\ }\bibfield  {title} {\enquote {\bibinfo {title}
  {{Self-stimulated pulse echo trains from inhomogeneously broadened spin
  ensembles}},}\ }\href {http://arxiv.org/abs/2004.01116v1} {\bibfield
  {journal} {\bibinfo  {journal} {arXiv}\ } (\bibinfo {year} {2020})},\ \Eprint
  {http://arxiv.org/abs/2004.01116v1} {2004.01116v1} \BibitemShut {NoStop}%
\bibitem [{\citenamefont {Weichselbaumer}\ \emph {et~al.}(2019)\citenamefont
  {Weichselbaumer}, \citenamefont {Natzkin}, \citenamefont {Zollitsch},
  \citenamefont {Weiler}, \citenamefont {Gross},\ and\ \citenamefont
  {Huebl}}]{Weichselbaumer2019}%
  \BibitemOpen
  \bibfield  {author} {\bibinfo {author} {\bibfnamefont {Stefan}\ \bibnamefont
  {Weichselbaumer}}, \bibinfo {author} {\bibfnamefont {Petio}\ \bibnamefont
  {Natzkin}}, \bibinfo {author} {\bibfnamefont {Christoph~W.}\ \bibnamefont
  {Zollitsch}}, \bibinfo {author} {\bibfnamefont {Mathias}\ \bibnamefont
  {Weiler}}, \bibinfo {author} {\bibfnamefont {Rudolf}\ \bibnamefont {Gross}},
  \ and\ \bibinfo {author} {\bibfnamefont {Hans}\ \bibnamefont {Huebl}},\
  }\bibfield  {title} {{\selectlanguage {english}\enquote {\bibinfo {title}
  {Quantitative {{Modeling}} of {{Superconducting Planar Resonators}} for
  {{Electron Spin Resonance}}},}\ }}\href {\doibase
  10.1103/PhysRevApplied.12.024021} {\bibfield  {journal} {\bibinfo  {journal}
  {Physical Review Applied}\ }\textbf {\bibinfo {volume} {12}},\ \bibinfo
  {pages} {024021} (\bibinfo {year} {2019})}\BibitemShut {NoStop}%
\bibitem [{CST(2016)}]{CST2016}%
  \BibitemOpen
  \href@noop {} {\enquote {\bibinfo {title} {{{CST Microwave Studio}} 2016},}\
  }\bibinfo {howpublished} {CST Computer Simulation Technology GmbH} (\bibinfo
  {year} {2016})\BibitemShut {NoStop}%
\bibitem [{\citenamefont {Wesenberg}\ \emph {et~al.}(2009)\citenamefont
  {Wesenberg}, \citenamefont {Ardavan}, \citenamefont {Briggs}, \citenamefont
  {Morton}, \citenamefont {Schoelkopf}, \citenamefont {Schuster},\ and\
  \citenamefont {M\o{}lmer}}]{Wesenberg2009}%
  \BibitemOpen
  \bibfield  {author} {\bibinfo {author} {\bibfnamefont {J.~H.}\ \bibnamefont
  {Wesenberg}}, \bibinfo {author} {\bibfnamefont {A.}~\bibnamefont {Ardavan}},
  \bibinfo {author} {\bibfnamefont {G.~A.~D.}\ \bibnamefont {Briggs}}, \bibinfo
  {author} {\bibfnamefont {J.~J.~L.}\ \bibnamefont {Morton}}, \bibinfo {author}
  {\bibfnamefont {R.~J.}\ \bibnamefont {Schoelkopf}}, \bibinfo {author}
  {\bibfnamefont {D.~I.}\ \bibnamefont {Schuster}}, \ and\ \bibinfo {author}
  {\bibfnamefont {K.}~\bibnamefont {M\o{}lmer}},\ }\bibfield  {title}
  {{\selectlanguage {english}\enquote {\bibinfo {title} {Quantum {{Computing}}
  with an {{Electron Spin Ensemble}}},}\ }}\href {\doibase
  10.1103/PhysRevLett.103.070502} {\bibfield  {journal} {\bibinfo  {journal}
  {Physical Review Letters}\ }\textbf {\bibinfo {volume} {103}},\ \bibinfo
  {pages} {070502} (\bibinfo {year} {2009})}\BibitemShut {NoStop}%
\bibitem [{\citenamefont {Feher}(1959)}]{Feher1959}%
  \BibitemOpen
  \bibfield  {author} {\bibinfo {author} {\bibfnamefont {G.}~\bibnamefont
  {Feher}},\ }\bibfield  {title} {\enquote {\bibinfo {title} {Electron spin
  resonance experiments on donors in silicon. {{I}}. {{Electronic}} structure
  of donors by the electron nuclear double resonance technique},}\ }\href
  {\doibase 10.1103/PhysRev.114.1219} {\bibfield  {journal} {\bibinfo
  {journal} {Physical Review}\ }\textbf {\bibinfo {volume} {114}},\ \bibinfo
  {pages} {1219} (\bibinfo {year} {1959})}\BibitemShut {NoStop}%
\bibitem [{\citenamefont {Schoelkopf}\ and\ \citenamefont
  {Girvin}(2008)}]{Schoelkopf2008}%
  \BibitemOpen
  \bibfield  {author} {\bibinfo {author} {\bibfnamefont {R.~J.}\ \bibnamefont
  {Schoelkopf}}\ and\ \bibinfo {author} {\bibfnamefont {S.~M.}\ \bibnamefont
  {Girvin}},\ }\bibfield  {title} {\enquote {\bibinfo {title} {Wiring up
  quantum systems},}\ }\href@noop {} {\bibfield  {journal} {\bibinfo  {journal}
  {Nature}\ }\textbf {\bibinfo {volume} {451}},\ \bibinfo {pages} {664--669}
  (\bibinfo {year} {2008})}\BibitemShut {NoStop}%
\bibitem [{\citenamefont {Chiorescu}\ \emph {et~al.}(2010)\citenamefont
  {Chiorescu}, \citenamefont {Groll}, \citenamefont {Bertaina}, \citenamefont
  {Mori},\ and\ \citenamefont {Miyashita}}]{Chiorescu:2010hw}%
  \BibitemOpen
  \bibfield  {author} {\bibinfo {author} {\bibfnamefont {I}~\bibnamefont
  {Chiorescu}}, \bibinfo {author} {\bibfnamefont {N}~\bibnamefont {Groll}},
  \bibinfo {author} {\bibfnamefont {S}~\bibnamefont {Bertaina}}, \bibinfo
  {author} {\bibfnamefont {T}~\bibnamefont {Mori}}, \ and\ \bibinfo {author}
  {\bibfnamefont {S}~\bibnamefont {Miyashita}},\ }\bibfield  {title}
  {{\selectlanguage {English}\enquote {\bibinfo {title} {{Magnetic strong
  coupling in a spin-photon system and transition to classical regime}},}\
  }}\href {\doibase 10.1103/PhysRevB.82.024413} {\bibfield  {journal} {\bibinfo
   {journal} {Phys Rev B}\ }\textbf {\bibinfo {volume} {82}},\ \bibinfo {pages}
  {024413} (\bibinfo {year} {2010})}\BibitemShut {NoStop}%
\bibitem [{\citenamefont {Bienfait}\ \emph {et~al.}(2016)\citenamefont
  {Bienfait}, \citenamefont {Pla}, \citenamefont {Kubo}, \citenamefont {Zhou},
  \citenamefont {Stern}, \citenamefont {Lo}, \citenamefont {Weis},
  \citenamefont {Schenkel}, \citenamefont {Vion}, \citenamefont {Esteve},
  \citenamefont {Morton},\ and\ \citenamefont {Bertet}}]{Bienfait2016a}%
  \BibitemOpen
  \bibfield  {author} {\bibinfo {author} {\bibfnamefont {A.}~\bibnamefont
  {Bienfait}}, \bibinfo {author} {\bibfnamefont {J.~J.}\ \bibnamefont {Pla}},
  \bibinfo {author} {\bibfnamefont {Y.}~\bibnamefont {Kubo}}, \bibinfo {author}
  {\bibfnamefont {X.}~\bibnamefont {Zhou}}, \bibinfo {author} {\bibfnamefont
  {M.}~\bibnamefont {Stern}}, \bibinfo {author} {\bibfnamefont {C.~C.}\
  \bibnamefont {Lo}}, \bibinfo {author} {\bibfnamefont {C.~D.}\ \bibnamefont
  {Weis}}, \bibinfo {author} {\bibfnamefont {T.}~\bibnamefont {Schenkel}},
  \bibinfo {author} {\bibfnamefont {D.}~\bibnamefont {Vion}}, \bibinfo {author}
  {\bibfnamefont {D.}~\bibnamefont {Esteve}}, \bibinfo {author} {\bibfnamefont
  {J.~J.~L.}\ \bibnamefont {Morton}}, \ and\ \bibinfo {author} {\bibfnamefont
  {P.}~\bibnamefont {Bertet}},\ }\bibfield  {title} {\enquote {\bibinfo {title}
  {Controlling spin relaxation with a cavity},}\ }\href {\doibase
  10.1038/nature16944} {\bibfield  {journal} {\bibinfo  {journal} {Nature}\
  }\textbf {\bibinfo {volume} {531}},\ \bibinfo {pages} {74--77} (\bibinfo
  {year} {2016})}\BibitemShut {NoStop}%
\bibitem [{\citenamefont {Klauder}\ and\ \citenamefont
  {Anderson}(1962)}]{Klauder1962}%
  \BibitemOpen
  \bibfield  {author} {\bibinfo {author} {\bibfnamefont {J.~R.}\ \bibnamefont
  {Klauder}}\ and\ \bibinfo {author} {\bibfnamefont {P.~W.}\ \bibnamefont
  {Anderson}},\ }\bibfield  {title} {{\selectlanguage {english}\enquote
  {\bibinfo {title} {Spectral {{Diffusion Decay}} in {{Spin Resonance
  Experiments}}},}\ }}\href {\doibase 10.1103/PhysRev.125.912} {\bibfield
  {journal} {\bibinfo  {journal} {Physical Review}\ }\textbf {\bibinfo {volume}
  {125}},\ \bibinfo {pages} {912} (\bibinfo {year} {1962})}\BibitemShut
  {NoStop}%
\bibitem [{\citenamefont {Tyryshkin}\ \emph {et~al.}(2003)\citenamefont
  {Tyryshkin}, \citenamefont {Lyon}, \citenamefont {Astashkin},\ and\
  \citenamefont {Raitsimring}}]{Tyryshkin2003}%
  \BibitemOpen
  \bibfield  {author} {\bibinfo {author} {\bibfnamefont {A.~M.}\ \bibnamefont
  {Tyryshkin}}, \bibinfo {author} {\bibfnamefont {S.~A.}\ \bibnamefont {Lyon}},
  \bibinfo {author} {\bibfnamefont {A.~V.}\ \bibnamefont {Astashkin}}, \ and\
  \bibinfo {author} {\bibfnamefont {A.~M.}\ \bibnamefont {Raitsimring}},\
  }\bibfield  {title} {{\selectlanguage {english}\enquote {\bibinfo {title}
  {Electron spin relaxation times of phosphorus donors in silicon},}\ }}\href
  {\doibase 10.1103/PhysRevB.68.193207} {\bibfield  {journal} {\bibinfo
  {journal} {Physical Review B}\ }\textbf {\bibinfo {volume} {68}},\ \bibinfo
  {pages} {193207} (\bibinfo {year} {2003})}\BibitemShut {NoStop}%
\bibitem [{\citenamefont {Shankar}\ \emph {et~al.}(2015)\citenamefont
  {Shankar}, \citenamefont {Tyryshkin},\ and\ \citenamefont
  {Lyon}}]{Shankar2015}%
  \BibitemOpen
  \bibfield  {author} {\bibinfo {author} {\bibfnamefont {S.}~\bibnamefont
  {Shankar}}, \bibinfo {author} {\bibfnamefont {A.~M.}\ \bibnamefont
  {Tyryshkin}}, \ and\ \bibinfo {author} {\bibfnamefont {S.~A.}\ \bibnamefont
  {Lyon}},\ }\bibfield  {title} {{\selectlanguage {english}\enquote {\bibinfo
  {title} {{{ESR}} measurements of phosphorus dimers in isotopically enriched
  $^{28}$si silicon},}\ }}\href {\doibase 10.1103/PhysRevB.91.245206}
  {\bibfield  {journal} {\bibinfo  {journal} {Physical Review B}\ }\textbf
  {\bibinfo {volume} {91}},\ \bibinfo {pages} {245206} (\bibinfo {year}
  {2015})}\BibitemShut {NoStop}%
\bibitem [{\citenamefont {Salikhov}\ \emph {et~al.}(1981)\citenamefont
  {Salikhov}, \citenamefont {Dzuba},\ and\ \citenamefont
  {Raitsimring}}]{Salikhov:1981ea}%
  \BibitemOpen
  \bibfield  {author} {\bibinfo {author} {\bibfnamefont {K~M}\ \bibnamefont
  {Salikhov}}, \bibinfo {author} {\bibfnamefont {S~A}\ \bibnamefont {Dzuba}}, \
  and\ \bibinfo {author} {\bibfnamefont {A~M}\ \bibnamefont {Raitsimring}},\
  }\bibfield  {title} {\enquote {\bibinfo {title} {{The theory of electron
  spin-echo signal decay resulting from dipole-dipole interactions between
  paramagnetic centers in solids}},}\ }\href {\doibase
  10.1016/0022-2364(81)90216-X} {\bibfield  {journal} {\bibinfo  {journal}
  {Journal of Magnetic Resonance}\ }\textbf {\bibinfo {volume} {42}},\ \bibinfo
  {pages} {255} (\bibinfo {year} {1981})}\BibitemShut {NoStop}%
\bibitem [{\citenamefont {Tyryshkin}\ \emph {et~al.}(2011)\citenamefont
  {Tyryshkin}, \citenamefont {Tojo}, \citenamefont {Morton}, \citenamefont
  {Riemann}, \citenamefont {Abrosimov}, \citenamefont {Becker}, \citenamefont
  {Pohl}, \citenamefont {Schenkel}, \citenamefont {Thewalt}, \citenamefont
  {Itoh},\ and\ \citenamefont {Lyon}}]{Tyryshkin:2011fi}%
  \BibitemOpen
  \bibfield  {author} {\bibinfo {author} {\bibfnamefont {Alexei~M}\
  \bibnamefont {Tyryshkin}}, \bibinfo {author} {\bibfnamefont {Shinichi}\
  \bibnamefont {Tojo}}, \bibinfo {author} {\bibfnamefont {John J~L}\
  \bibnamefont {Morton}}, \bibinfo {author} {\bibfnamefont {Helge}\
  \bibnamefont {Riemann}}, \bibinfo {author} {\bibfnamefont {Nikolai~V}\
  \bibnamefont {Abrosimov}}, \bibinfo {author} {\bibfnamefont {Peter}\
  \bibnamefont {Becker}}, \bibinfo {author} {\bibfnamefont {Hans-Joachim}\
  \bibnamefont {Pohl}}, \bibinfo {author} {\bibfnamefont {Thomas}\ \bibnamefont
  {Schenkel}}, \bibinfo {author} {\bibfnamefont {Michael L~W}\ \bibnamefont
  {Thewalt}}, \bibinfo {author} {\bibfnamefont {Kohei~M}\ \bibnamefont {Itoh}},
  \ and\ \bibinfo {author} {\bibfnamefont {S~A}\ \bibnamefont {Lyon}},\
  }\bibfield  {title} {\enquote {\bibinfo {title} {{Electron spin coherence
  exceeding seconds in high-purity silicon}},}\ }\href {\doibase
  10.1038/nmat3182} {\bibfield  {journal} {\bibinfo  {journal} {Nature
  Materials}\ }\textbf {\bibinfo {volume} {11}},\ \bibinfo {pages} {143}
  (\bibinfo {year} {2011})}\BibitemShut {NoStop}%
\bibitem [{\citenamefont {Taylor}\ \emph {et~al.}(1974)\citenamefont {Taylor},
  \citenamefont {Marko},\ and\ \citenamefont {Bartlet}}]{Taylor:1974}%
  \BibitemOpen
  \bibfield  {author} {\bibinfo {author} {\bibfnamefont {D~R}\ \bibnamefont
  {Taylor}}, \bibinfo {author} {\bibfnamefont {J~R}\ \bibnamefont {Marko}}, \
  and\ \bibinfo {author} {\bibfnamefont {I~G}\ \bibnamefont {Bartlet}},\
  }\bibfield  {title} {\enquote {\bibinfo {title} {{Exchange and Dipolar FIelds
  in Phosphorus-Doped Silicon Measured by Electron Spin Resonance Echoes}},}\
  }\href@noop {} {\bibfield  {journal} {\bibinfo  {journal} {Solid State
  Commun}\ }\textbf {\bibinfo {volume} {14}},\ \bibinfo {pages} {295} (\bibinfo
  {year} {1974})}\BibitemShut {NoStop}%
\bibitem [{\citenamefont {Alaimo}\ and\ \citenamefont
  {Roberts}(1997)}]{Alaimo1997}%
  \BibitemOpen
  \bibfield  {author} {\bibinfo {author} {\bibfnamefont {M.H.}\ \bibnamefont
  {Alaimo}}\ and\ \bibinfo {author} {\bibfnamefont {J.E.}\ \bibnamefont
  {Roberts}},\ }\bibfield  {title} {{\selectlanguage {english}\enquote
  {\bibinfo {title} {Effects of paramagnetic cations on the nonexponential
  spin-lattice relaxation of rare spin nuclei in solids},}\ }}\href {\doibase
  10.1016/S0926-2040(97)00006-4} {\bibfield  {journal} {\bibinfo  {journal}
  {Solid State Nuclear Magnetic Resonance}\ }\textbf {\bibinfo {volume} {8}},\
  \bibinfo {pages} {241--250} (\bibinfo {year} {1997})}\BibitemShut {NoStop}%
\bibitem [{\citenamefont {Bloembergen}(1949)}]{Bloembergen:1949gv}%
  \BibitemOpen
  \bibfield  {author} {\bibinfo {author} {\bibfnamefont {N}~\bibnamefont
  {Bloembergen}},\ }\bibfield  {title} {{\selectlanguage {English}\enquote
  {\bibinfo {title} {{On the interaction of nuclear spins in a crystalline
  lattice}},}\ }}\href {\doibase 10.1016/0031-8914(49)90114-7} {\bibfield
  {journal} {\bibinfo  {journal} {Physica}\ }\textbf {\bibinfo {volume} {15}},\
  \bibinfo {pages} {386--426} (\bibinfo {year} {1949})}\BibitemShut {NoStop}%
\bibitem [{\citenamefont {Eberhardt}\ \emph {et~al.}(2007)\citenamefont
  {Eberhardt}, \citenamefont {Mouaziz}, \citenamefont {Boero}, \citenamefont
  {Brugger},\ and\ \citenamefont {Meier}}]{Eberhardt:2007fh}%
  \BibitemOpen
  \bibfield  {author} {\bibinfo {author} {\bibfnamefont {Kai~W}\ \bibnamefont
  {Eberhardt}}, \bibinfo {author} {\bibfnamefont {Schahrazede}\ \bibnamefont
  {Mouaziz}}, \bibinfo {author} {\bibfnamefont {Giovanni}\ \bibnamefont
  {Boero}}, \bibinfo {author} {\bibfnamefont {J{\"u}rgen}\ \bibnamefont
  {Brugger}}, \ and\ \bibinfo {author} {\bibfnamefont {Beat~H}\ \bibnamefont
  {Meier}},\ }\bibfield  {title} {{\selectlanguage {English}\enquote {\bibinfo
  {title} {{Direct Observation of Nuclear Spin Diffusion in Real Space}},}\
  }}\href {\doibase 10.1103/PhysRevLett.99.227603} {\bibfield  {journal}
  {\bibinfo  {journal} {Physical Review Letters}\ }\textbf {\bibinfo {volume}
  {99}},\ \bibinfo {pages} {227603} (\bibinfo {year} {2007})}\BibitemShut
  {NoStop}%
\bibitem [{\citenamefont {Redfield}(1959)}]{Redfield:1959hc}%
  \BibitemOpen
  \bibfield  {author} {\bibinfo {author} {\bibfnamefont {A~G}\ \bibnamefont
  {Redfield}},\ }\bibfield  {title} {{\selectlanguage {English}\enquote
  {\bibinfo {title} {{Spatial Diffusion of Spin Energy}},}\ }}\href {\doibase
  10.1103/PhysRev.116.315} {\bibfield  {journal} {\bibinfo  {journal} {Physical
  Review}\ }\textbf {\bibinfo {volume} {116}},\ \bibinfo {pages} {315--316}
  (\bibinfo {year} {1959})}\BibitemShut {NoStop}%
\bibitem [{\citenamefont {Vugmeister}(1976)}]{Vugmeister:1976co}%
  \BibitemOpen
  \bibfield  {author} {\bibinfo {author} {\bibfnamefont {B~E}\ \bibnamefont
  {Vugmeister}},\ }\bibfield  {title} {\enquote {\bibinfo {title} {{Spatial and
  Spectral Spin Diffusion in Dilute Spin Systems}},}\ }\href {\doibase
  10.1002/pssb.2220760116} {\bibfield  {journal} {\bibinfo  {journal} {physica
  status solidi (b)}\ }\textbf {\bibinfo {volume} {76}},\ \bibinfo {pages}
  {161--170} (\bibinfo {year} {1976})}\BibitemShut {NoStop}%
\bibitem [{\citenamefont {Feher}\ and\ \citenamefont
  {Gere}(1959)}]{Feher1959b}%
  \BibitemOpen
  \bibfield  {author} {\bibinfo {author} {\bibfnamefont {G}~\bibnamefont
  {Feher}}\ and\ \bibinfo {author} {\bibfnamefont {E~A}\ \bibnamefont {Gere}},\
  }\bibfield  {title} {{\selectlanguage {English}\enquote {\bibinfo {title}
  {{Electron Spin Resonance Experiments on Donors in Silicon. II. Electron Spin
  Relaxation Effects}},}\ }}\href {\doibase 10.1103/PhysRev.114.1245}
  {\bibfield  {journal} {\bibinfo  {journal} {Physical Review}\ }\textbf
  {\bibinfo {volume} {114}},\ \bibinfo {pages} {1245} (\bibinfo {year}
  {1959})}\BibitemShut {NoStop}%
\bibitem [{\citenamefont {Morello}\ \emph {et~al.}(2010)\citenamefont
  {Morello}, \citenamefont {Pla}, \citenamefont {Zwanenburg}, \citenamefont
  {Chan}, \citenamefont {Tan}, \citenamefont {Huebl}, \citenamefont {Mottonen},
  \citenamefont {Nugroho}, \citenamefont {Yang}, \citenamefont {van Donkelaar},
  \citenamefont {Alves}, \citenamefont {Jamieson}, \citenamefont {Escott},
  \citenamefont {Hollenberg}, \citenamefont {Clark},\ and\ \citenamefont
  {Dzurak}}]{morello2010}%
  \BibitemOpen
  \bibfield  {author} {\bibinfo {author} {\bibfnamefont {Andrea}\ \bibnamefont
  {Morello}}, \bibinfo {author} {\bibfnamefont {Jarryd~J}\ \bibnamefont {Pla}},
  \bibinfo {author} {\bibfnamefont {Floris~A}\ \bibnamefont {Zwanenburg}},
  \bibinfo {author} {\bibfnamefont {Kok~W}\ \bibnamefont {Chan}}, \bibinfo
  {author} {\bibfnamefont {Kuan~Y}\ \bibnamefont {Tan}}, \bibinfo {author}
  {\bibfnamefont {Hans}\ \bibnamefont {Huebl}}, \bibinfo {author}
  {\bibfnamefont {Mikko}\ \bibnamefont {Mottonen}}, \bibinfo {author}
  {\bibfnamefont {Christopher~D}\ \bibnamefont {Nugroho}}, \bibinfo {author}
  {\bibfnamefont {Changyi}\ \bibnamefont {Yang}}, \bibinfo {author}
  {\bibfnamefont {Jessica~A}\ \bibnamefont {van Donkelaar}}, \bibinfo {author}
  {\bibfnamefont {Andrew D~C}\ \bibnamefont {Alves}}, \bibinfo {author}
  {\bibfnamefont {David~N}\ \bibnamefont {Jamieson}}, \bibinfo {author}
  {\bibfnamefont {Christopher~C}\ \bibnamefont {Escott}}, \bibinfo {author}
  {\bibfnamefont {Lloyd C~L}\ \bibnamefont {Hollenberg}}, \bibinfo {author}
  {\bibfnamefont {Robert~G}\ \bibnamefont {Clark}}, \ and\ \bibinfo {author}
  {\bibfnamefont {Andrew~S}\ \bibnamefont {Dzurak}},\ }\bibfield  {title}
  {\enquote {\bibinfo {title} {{Single-shot readout of an electron spin in
  silicon}},}\ }\href {\doibase 10.1038/nature09392} {\bibfield  {journal}
  {\bibinfo  {journal} {Nature}\ }\textbf {\bibinfo {volume} {467}},\ \bibinfo
  {pages} {687} (\bibinfo {year} {2010})}\BibitemShut {NoStop}%
\bibitem [{\citenamefont {Hasegawa}(1960)}]{Hasegawa:1960ey}%
  \BibitemOpen
  \bibfield  {author} {\bibinfo {author} {\bibfnamefont {Hiroshi}\ \bibnamefont
  {Hasegawa}},\ }\bibfield  {title} {{\selectlanguage {English}\enquote
  {\bibinfo {title} {{Spin-Lattice Relaxation of Shallow Donor States in Ge and
  Si through a Direct Phonon Process}},}\ }}\href {\doibase
  10.1103/PhysRev.118.1523} {\bibfield  {journal} {\bibinfo  {journal}
  {Physical Review}\ }\textbf {\bibinfo {volume} {118}},\ \bibinfo {pages}
  {1523} (\bibinfo {year} {1960})}\BibitemShut {NoStop}%
\bibitem [{\citenamefont {Tavis}\ and\ \citenamefont
  {Cummings}(1968)}]{Tavis1968a}%
  \BibitemOpen
  \bibfield  {author} {\bibinfo {author} {\bibfnamefont {M.}~\bibnamefont
  {Tavis}}\ and\ \bibinfo {author} {\bibfnamefont {F.W.}\ \bibnamefont
  {Cummings}},\ }\bibfield  {title} {\enquote {\bibinfo {title} {{Exact
  Solution for an N-Molecule-Radiation-Field Hamiltonian}},}\ }\href
  {https://link.aps.org/doi/10.1103/PhysRev.170.379} {\bibfield  {journal}
  {\bibinfo  {journal} {Physical Review}\ }\textbf {\bibinfo {volume} {170}},\
  \bibinfo {pages} {379--384} (\bibinfo {year} {1968})}\BibitemShut {NoStop}%
\bibitem [{\citenamefont {Carmichael}(2007)}]{CarmichaelQO2}%
  \BibitemOpen
  \bibfield  {author} {\bibinfo {author} {\bibfnamefont {H~J}\ \bibnamefont
  {Carmichael}},\ }\href {https://books.google.at/books?id=xgxOYkxW8JoC} {
  {\bibinfo {title} {{Statistical Methods in Quantum Optics 2: Non-Classical
  Fields}}}},\ Theoretical and Mathematical Physics\ (\bibinfo  {publisher}
  {Springer Berlin Heidelberg},\ \bibinfo {year} {2007})\BibitemShut {NoStop}%
\bibitem [{\citenamefont {Sandner}\ \emph {et~al.}(2012)\citenamefont
  {Sandner}, \citenamefont {Ritsch}, \citenamefont {Ams\"uss}, \citenamefont
  {Koller}, \citenamefont {N\"obauer}, \citenamefont {Putz}, \citenamefont
  {Schmiedmayer},\ and\ \citenamefont {Majer}}]{Sandner2012}%
  \BibitemOpen
  \bibfield  {author} {\bibinfo {author} {\bibfnamefont {K.}~\bibnamefont
  {Sandner}}, \bibinfo {author} {\bibfnamefont {H.}~\bibnamefont {Ritsch}},
  \bibinfo {author} {\bibfnamefont {R.}~\bibnamefont {Ams\"uss}}, \bibinfo
  {author} {\bibfnamefont {Ch.}\ \bibnamefont {Koller}}, \bibinfo {author}
  {\bibfnamefont {T.}~\bibnamefont {N\"obauer}}, \bibinfo {author}
  {\bibfnamefont {S.}~\bibnamefont {Putz}}, \bibinfo {author} {\bibfnamefont
  {J.}~\bibnamefont {Schmiedmayer}}, \ and\ \bibinfo {author} {\bibfnamefont
  {J.}~\bibnamefont {Majer}},\ }\bibfield  {title} {{\selectlanguage
  {english}\enquote {\bibinfo {title} {Strong magnetic coupling of an
  inhomogeneous nitrogen-vacancy ensemble to a cavity},}\ }}\href {\doibase
  10.1103/PhysRevA.85.053806} {\bibfield  {journal} {\bibinfo  {journal}
  {Physical Review A}\ }\textbf {\bibinfo {volume} {85}},\ \bibinfo {pages}
  {053806} (\bibinfo {year} {2012})}\BibitemShut {NoStop}%
\bibitem [{\citenamefont {Deville}\ \emph {et~al.}(1979)\citenamefont
  {Deville}, \citenamefont {Bernier},\ and\ \citenamefont
  {Delrieux}}]{Deville1979}%
  \BibitemOpen
  \bibfield  {author} {\bibinfo {author} {\bibfnamefont {G.}~\bibnamefont
  {Deville}}, \bibinfo {author} {\bibfnamefont {M.}~\bibnamefont {Bernier}}, \
  and\ \bibinfo {author} {\bibfnamefont {J.~M.}\ \bibnamefont {Delrieux}},\
  }\bibfield  {title} {{\selectlanguage {english}\enquote {\bibinfo {title}
  {{{NMR}} multiple echoes observed in solid $^3${{He}}},}\ }}\href {\doibase
  10.1103/PhysRevB.19.5666} {\bibfield  {journal} {\bibinfo  {journal}
  {Physical Review B}\ }\textbf {\bibinfo {volume} {19}},\ \bibinfo {pages}
  {5666} (\bibinfo {year} {1979})}\BibitemShut {NoStop}%
\bibitem [{\citenamefont {Eska}\ \emph {et~al.}(1981)\citenamefont {Eska},
  \citenamefont {Willers}, \citenamefont {Amend},\ and\ \citenamefont
  {Wiedemann}}]{Eska1981}%
  \BibitemOpen
  \bibfield  {author} {\bibinfo {author} {\bibfnamefont {G.}~\bibnamefont
  {Eska}}, \bibinfo {author} {\bibfnamefont {H.-G.}\ \bibnamefont {Willers}},
  \bibinfo {author} {\bibfnamefont {B.}~\bibnamefont {Amend}}, \ and\ \bibinfo
  {author} {\bibfnamefont {W.}~\bibnamefont {Wiedemann}},\ }\bibfield  {title}
  {{\selectlanguage {english}\enquote {\bibinfo {title} {Spin echo experiments
  in superfluid {{3He}}},}\ }}\href {\doibase 10.1016/0378-4363(81)90878-0}
  {\bibfield  {journal} {\bibinfo  {journal} {Physica B+C}\ }\textbf {\bibinfo
  {volume} {108}},\ \bibinfo {pages} {1155} (\bibinfo {year}
  {1981})}\BibitemShut {NoStop}%
\bibitem [{\citenamefont {Einzel}\ \emph {et~al.}(1984)\citenamefont {Einzel},
  \citenamefont {Eska}, \citenamefont {Hirayoshi}, \citenamefont {Kopp},\ and\
  \citenamefont {Wolfle}}]{Einzel1984}%
  \BibitemOpen
  \bibfield  {author} {\bibinfo {author} {\bibfnamefont {D}~\bibnamefont
  {Einzel}}, \bibinfo {author} {\bibfnamefont {G}~\bibnamefont {Eska}},
  \bibinfo {author} {\bibfnamefont {Y}~\bibnamefont {Hirayoshi}}, \bibinfo
  {author} {\bibfnamefont {T}~\bibnamefont {Kopp}}, \ and\ \bibinfo {author}
  {\bibfnamefont {P}~\bibnamefont {Wolfle}},\ }\bibfield  {title}
  {{\selectlanguage {english}\enquote {\bibinfo {title} {Multiple {{Spin
  Echoes}} in a {{Normal Fermi Liquid}}},}\ }}\href {\doibase
  10.1103/PhysRevLett.53.2312} {\bibfield  {journal} {\bibinfo  {journal}
  {Physical Review Letters}\ }\textbf {\bibinfo {volume} {53}},\ \bibinfo
  {pages} {2312} (\bibinfo {year} {1984})}\BibitemShut {NoStop}%
\bibitem [{\citenamefont {Bowtell}\ \emph {et~al.}(1990)\citenamefont
  {Bowtell}, \citenamefont {Bowley},\ and\ \citenamefont
  {Glover}}]{Bowtell1990}%
  \BibitemOpen
  \bibfield  {author} {\bibinfo {author} {\bibfnamefont {R}~\bibnamefont
  {Bowtell}}, \bibinfo {author} {\bibfnamefont {R.M}\ \bibnamefont {Bowley}}, \
  and\ \bibinfo {author} {\bibfnamefont {P}~\bibnamefont {Glover}},\ }\bibfield
   {title} {{\selectlanguage {english}\enquote {\bibinfo {title} {Multiple spin
  echoes in liquids in a high magnetic field},}\ }}\href {\doibase
  10.1016/0022-2364(90)90297-M} {\bibfield  {journal} {\bibinfo  {journal}
  {Journal of Magnetic Resonance}\ }\textbf {\bibinfo {volume} {88}},\ \bibinfo
  {pages} {643} (\bibinfo {year} {1990})}\BibitemShut {NoStop}%
\bibitem [{\citenamefont {Bedford}\ \emph {et~al.}(1991)\citenamefont
  {Bedford}, \citenamefont {Bowley}, \citenamefont {{Owers-Bradley}},\ and\
  \citenamefont {Wightman}}]{Bedford1991a}%
  \BibitemOpen
  \bibfield  {author} {\bibinfo {author} {\bibfnamefont {A.~S.}\ \bibnamefont
  {Bedford}}, \bibinfo {author} {\bibfnamefont {R.~M.}\ \bibnamefont {Bowley}},
  \bibinfo {author} {\bibfnamefont {J.~R.}\ \bibnamefont {{Owers-Bradley}}}, \
  and\ \bibinfo {author} {\bibfnamefont {D.}~\bibnamefont {Wightman}},\
  }\bibfield  {title} {{\selectlanguage {english}\enquote {\bibinfo {title}
  {Multiple spin echoes in spin polarized {{Fermi}} liquids},}\ }}\href
  {\doibase 10.1007/BF00682194} {\bibfield  {journal} {\bibinfo  {journal}
  {Journal of Low Temperature Physics}\ }\textbf {\bibinfo {volume} {85}},\
  \bibinfo {pages} {389} (\bibinfo {year} {1991})}\BibitemShut {NoStop}%
\bibitem [{\citenamefont {Bowtell}\ and\ \citenamefont
  {Robyr}(1996)}]{Bowtell1996}%
  \BibitemOpen
  \bibfield  {author} {\bibinfo {author} {\bibfnamefont {R.}~\bibnamefont
  {Bowtell}}\ and\ \bibinfo {author} {\bibfnamefont {P.}~\bibnamefont
  {Robyr}},\ }\bibfield  {title} {{\selectlanguage {english}\enquote {\bibinfo
  {title} {Structural {{Investigations}} with the {{Dipolar Demagnetizing
  Field}} in {{Solution NMR}}},}\ }}\href {\doibase
  10.1103/PhysRevLett.76.4971} {\bibfield  {journal} {\bibinfo  {journal}
  {Physical Review Letters}\ }\textbf {\bibinfo {volume} {76}},\ \bibinfo
  {pages} {4971} (\bibinfo {year} {1996})}\BibitemShut {NoStop}%
\bibitem [{\citenamefont {Warren}\ \emph {et~al.}(1995)\citenamefont {Warren},
  \citenamefont {Lee}, \citenamefont {Richter},\ and\ \citenamefont
  {Vathyam}}]{Warren1995}%
  \BibitemOpen
  \bibfield  {author} {\bibinfo {author} {\bibfnamefont {W.S.}\ \bibnamefont
  {Warren}}, \bibinfo {author} {\bibfnamefont {S.}~\bibnamefont {Lee}},
  \bibinfo {author} {\bibfnamefont {W.}~\bibnamefont {Richter}}, \ and\
  \bibinfo {author} {\bibfnamefont {S.}~\bibnamefont {Vathyam}},\ }\bibfield
  {title} {{\selectlanguage {english}\enquote {\bibinfo {title} {Correcting the
  classical dipolar demagnetizing field in solution {{NMR}}},}\ }}\href
  {\doibase 10.1016/0009-2614(95)01184-5} {\bibfield  {journal} {\bibinfo
  {journal} {Chemical Physics Letters}\ }\textbf {\bibinfo {volume} {247}},\
  \bibinfo {pages} {207} (\bibinfo {year} {1995})}\BibitemShut {NoStop}%
\bibitem [{\citenamefont {Bloom}(1957)}]{Bloom1957}%
  \BibitemOpen
  \bibfield  {author} {\bibinfo {author} {\bibfnamefont {Stanley}\ \bibnamefont
  {Bloom}},\ }\bibfield  {title} {{\selectlanguage {english}\enquote {\bibinfo
  {title} {Effects of {{Radiation Damping}} on {{Spin Dynamics}}},}\ }}\href
  {\doibase 10.1063/1.1722859} {\bibfield  {journal} {\bibinfo  {journal}
  {Journal of Applied Physics}\ }\textbf {\bibinfo {volume} {28}},\ \bibinfo
  {pages} {800--805} (\bibinfo {year} {1957})}\BibitemShut {NoStop}%
\bibitem [{\citenamefont {Augustine}(2002)}]{Augustine2002}%
  \BibitemOpen
  \bibfield  {author} {\bibinfo {author} {\bibfnamefont {M.P.}\ \bibnamefont
  {Augustine}},\ }\bibfield  {title} {{\selectlanguage {english}\enquote
  {\bibinfo {title} {Transient properties of radiation damping},}\ }}\href
  {\doibase 10.1016/S0079-6565(01)00037-1} {\bibfield  {journal} {\bibinfo
  {journal} {Progress in Nuclear Magnetic Resonance Spectroscopy}\ }\textbf
  {\bibinfo {volume} {40}},\ \bibinfo {pages} {111--150} (\bibinfo {year}
  {2002})}\BibitemShut {NoStop}%
\bibitem [{\citenamefont {Vlassenbroek}\ \emph {et~al.}(1995)\citenamefont
  {Vlassenbroek}, \citenamefont {Jeener},\ and\ \citenamefont
  {Broekaert}}]{Vlassenbroek1995}%
  \BibitemOpen
  \bibfield  {author} {\bibinfo {author} {\bibfnamefont {A.}~\bibnamefont
  {Vlassenbroek}}, \bibinfo {author} {\bibfnamefont {J.}~\bibnamefont
  {Jeener}}, \ and\ \bibinfo {author} {\bibfnamefont {P.}~\bibnamefont
  {Broekaert}},\ }\bibfield  {title} {{\selectlanguage {english}\enquote
  {\bibinfo {title} {Radiation damping in high resolution liquid {{NMR}}: {{A}}
  simulation study},}\ }}\href {\doibase 10.1063/1.470468} {\bibfield
  {journal} {\bibinfo  {journal} {The Journal of Chemical Physics}\ }\textbf
  {\bibinfo {volume} {103}},\ \bibinfo {pages} {5886--5897} (\bibinfo {year}
  {1995})}\BibitemShut {NoStop}%
\end{thebibliography}
\end{document}